\documentclass[runningheads,twoside,12pt]{llncsfp} \def\versy{full}
\def\fully{full}

\usepackage{amssymb}
\usepackage[british,english]{babel}
\usepackage{amssymb,enumerate,multirow,makecell}
\usepackage{amsfonts,amsmath,breakcites}
\usepackage{latexsym,url}
\usepackage{microtype}
\spnewtheorem{thm}{Theorem}{\bf }{\it }
\spnewtheorem{prob}[thm]{Problem}{\bf }{\it }
\spnewtheorem{defn}[thm]{Definition}{\bf }{\rm }
\spnewtheorem{desc}[thm]{Description}{\bf }{\rm }
\spnewtheorem{prop}[thm]{Proposition}{\bf }{\it }
\spnewtheorem{cor}[thm]{Corollary}{\bf }{\it }
\spnewtheorem{rem}[thm]{Remark}{\bf }{\rm }
\spnewtheorem{clm}[thm]{Claim}{\bf }{\it }
\spnewtheorem{quest}[thm]{Question}{\bf }{\it }
\spnewtheorem{exmp}[thm]{Example}{\bf }{\rm }
\spnewtheorem{constr}[thm]{Construction}{\bf }{\rm }
\def\sp{\\* \indent}

\def\leq{\leqslant}
\def\geq{\geqslant}

\def\ownproof{\noindent{\bf Proof.}\ }

\def\setminus{-}

\DeclareMathOperator{\rng}{range} 
\DeclareMathOperator{\cnt}{content} 

\newcommand{\cP}{{\mathcal{P}}}

\newcommand{\cA}{{\mathcal{A}}}

\newcommand{\cR}{{\mathcal{R}}}
\newcommand\spn[1]{{\left\langle#1\right\rangle}}

\newcommand{\N}{\mathbb{N}}

\newcommand{\sm}{\setminus}

\newcommand{\cL}{\mathcal{L}}

\newcommand{\cF}{{\mathcal F}}
\newcommand{\cH}{\mathcal{H}}
\newcommand{\cC}{\mathcal{L}}

\newcommand{\BC}{\mbox{BC}}
\newcommand{\Vac}{\mbox{Vac}}
\newcommand{\Ex}{\mbox{Ex}}

\newcommand{\Conf}{\mbox{Conf}}

\newcommand{\Fin}{\mbox{Fin}}

\newcommand{\conv}{\!\downarrow}

\newcommand{\f}{\varphi}

\newcommand{\K}{K}

\begin{document}

\title{Learnability and Positive Equivalence Relations\thanks{D.~Belanger
(as RF), Z.~Gao (as RF) and S.~Jain (as Co-PI), F.~Stephan (as PI)
have been supported by the Singapore Ministry
of Education Academic Research Fund grant MOE2016-T2-1-019 / R146-000-234-112;
the latter three have also been suported by the Singapore Ministry
of Education Academic Research Fund grant MOE2019-T2-2-121 / R146-000-304-112.
Furthermore, S.~Jain is supported in part by NUS grant C252-000-087-001.
Part of the work was done while David Belanger worked at Ghent University
where he was supported by the BOF grant 01P01117.}}

\author{David Belanger\inst{1}, Ziyuan Gao\inst{2},
Sanjay Jain\inst{3}, \newline Wei Li\inst{2}
and Frank Stephan\inst{2}$^,$\inst{3}}

\authorrunning{D.~Belanger, Z.~Gao, S.~Jain, W.~Li and F.~Stephan}

\institute{Institute of Mathematical Sciences, National University of
Singapore \\  3 Prince George's Park, Singapore 118402, Republic of Singapore \\
\email{belanger@nus.edu.sg}
\and
Department of Mathematics, National University of Singapore \\
10 Lower Kent Ridge Road, Singapore 119076, Republic of Singapore \\
\email{\{matgaoz,matliw\}@nus.edu.sg}
\and
School of Computing, National University of
Singapore \\ Singapore 117417, Republic of Singapore \\
\email{\{sanjay,fstephan\}@comp.nus.edu.sg}}

\maketitle

\begin{abstract}
\noindent
Prior work of Gavryushkin, Khoussainov, Jain and Stephan investigated
what algebraic structures can be realised in worlds given by a
positive (= recursively enumerable) equivalence relation which
partitions the natural numbers into infinitely many equivalence
classes. The present work investigates the infinite one-one numbered
recursively enumerable (r.e.) families realised by such relations and
asks how the choice of the equivalence relation impacts the
learnability properties of these classes when studying learnability in
the limit from positive examples, also known as learning from text. 
For all choices of such positive equivalence relations, for each of
the following entries, there are one-one numbered r.e.\ families which
satisfy it: (a) they are behaviourally correctly learnable but not
vacillatorily learnable; (b) they are explanatorily learnable but not
confidently learnable; (c) they are not behaviourally correctly
learnable. Furthermore, there is a positive equivalence relation which
enforces that (d) every vacillatorily learnable one-one numbered
family of languages closed under this equivalence relation is already
explanatorily learnable and cannot be confidently learnable.
\end{abstract}

\section{Introduction}

\noindent
Consider a learning scenario where all positive examples of a given
target concept $L$ belonging to a concept class $\cC$ are shown
sequentially to a learner $M$.  After processing each example, $M$
makes a conjecture as to the identity of the target concept, based on
some fixed representation system of all concepts in $\cC$.  $M$ is
said to successfully identify $L$ if its sequence of conjectures
converges to a correct hypothesis
describing $L$.  This learning paradigm, due to Gold \cite{Go67}, is
well-studied and has inspired
the development of a large number of other learning models in
inductive inference \cite{JORS99,OSW86}.

In this work, we study how the interrelations between the elements of
a domain $X$ influences the learnability of classes of languages
defined over $X$.  The domain of interest throughout this work is
$\mathbb N$.  We will be concerned with \emph{recursively enumerable
(r.e.)} equivalence relations defined on $\mathbb N$ that induce {\em
infinitely} many equivalence classes.  The main motivation for
focussing on such relations comes from the study of r.e.\ structures. 
Here, r.e.\ structures are given by a domain, recursive functions
representing basic operators in the structure, and some recursively
enumerable predicates, among which there is a recursively
enumerable equivalence relation $\eta$ with infinitely many
equivalence classes which plays the role of equality in the given
structure.  Such structures have been studied for a long time; for
example, Novikov \cite{No55}
constructed a finitely generated group with undecidable word-problem;
in other words, there is a group which can be represented using an
r.e.\ but nonrecursive equivalence relation (as equality of the group)
but one cannot represent it using a recursive equivalence relation
$E$. On the other hand, for Noetherian rings \cite{Noe21}, Baur
\cite{Bau74} showed that every r.e.\ Noetherian ring is a recursive
ring, implying that the underlying equality $\eta$ is always a
recursive relation and that its equivalence classes are uniformly
recursive.  Another example of an r.e.\ equivalence relation is the
relation of \emph{provable equivalence} with respect to any formal
system, say Peano Arithmetic (PA), where
\begin{quote}
   $x \sim_{\text{PA}} y$ holds iff $\alpha \leftrightarrow \beta$
   is provable in PA.
\end{quote}
Here $x$ and $y$ are the G\"odel numbers of
$\alpha$ and $\beta$, respectively,
according to some fixed G\"odel numbering.  
\sp
Fokina, Gavryushkin, Jain, Khoussainov, Semukhin, Stephan and Turetsky
\cite{FKST16,GJKS14,GKS16} focussed in a sequence of papers on the
question of which type of structures could be realised by an r.e.\
equivalence relation on $\mathbb N$ with infinitely many equivalence
classes and how different relations compare with respect to their
ability to realise structures of certain type.  Ershov
\cite{Er74,Er77}, and following him Odifreddi \cite{Od89}, call r.e.\
equivalence relations {\em positive} equivalence relations.  Fokina,
K{\"o}tzing and San Mauro \cite{FKSM19} studied Gold-style
learnability of equivalence structures (with no computability
restrictions on the learner), and gave a structural characterisation
of families of equivalence structures with no infinite equivalence
classes that are learnable in the limit from informant.  In follow-up
work by Bazhenov, Fokina and San Mauro \cite{Bazhenov20}, a
model-theoretic characterisation of families of general equivalence
structures that are learnable in the limit from informant was obtained.

Now a structure is realised by a positive equivalence relation $\eta$
iff there is a bijection between the elements in the domain of the
structure and the equivalence classes of $\eta$ and all relations
involved are r.e.\ and all functions are realised by recursive
functions which respect $\eta$. In the simplest case of functions from
the domain to the domain, they respect $\eta$ if they map
$\eta$-equivalent numbers to $\eta$-equivalent numbers. This work
focusses on the study of learnability within the framework of families
realised by positive equivalence relations. In particular the topic of
the investigations is to what extent separations between learning
criteria known from inductive inference can be witnessed by
$\eta$-closed sets, that is, which of the
positive equivalence relations witness a separation of two learning
criteria or collapse them. Furthermore one asks, whether certain
learning criteria can be void (non-existent) for certain equivalence
relations $\eta$.
It could be shown that one single equivalence relation $\vartheta$
witnesses the two above mentioned collapses.
The study of learning in a world given by some $\eta$ is similar to
that in complexity relative to an oracle; one wants to know how robust
the results from the non-relativised world are and how much they generalise.

As mentioned earlier, a major topic investigated in this work is the
relationship between various learning criteria for a fixed but
arbitrary positive equivalence relation $\eta$ on $\mathbb N$.  In
particular, we study the question of whether the strict hierarchy of
learnability notions
$$
\Fin \subset \Conf \subset \Ex \subset \Vac \subset \BC
$$
(see Definition \ref{defnmain}), which holds with respect to the
general class of uniformly r.e.\ families (see \cite{JORS99}), carries
over to the class of $\eta$-families for any given positive
equivalence relation $\eta$. Here for any learning criteria $A$ and
$B$, we write $A \subset B$ to mean that the class of uniformly r.e.\
families that are learnable under criterion $A$ is a strict subset of
the class of uniformly r.e.\ families that
are learnable under criterion $B$.  The above inclusions follow
directly from the definitions of the respective learning criteria, so
the main interesting question is whether the inclusions are strict. 
It will be shown that many learnability relations that hold when
$\eta$ is just the usual equality relation on $\mathbb N$ carry over
for general $\eta$.  In particular, the fact that the family of
explanatorily learnable classes of languages is a strict subset of the
family of behaviourally correctly learnable classes of languages,
which is well-known when $\eta$ is the equality relation \cite[Theorem
3.1]{CS83}, holds for \emph{any} positive equivalence relation $\eta$
(see Theorem \ref{th:bcnotex}).  On the other hand, we show that for a
special positive equivalence relation 
$\vartheta$, every vacillatorily learnable $\vartheta$-family is
already explanatorily learnable (see Theorem
\ref{thm:varthetaproperties}); this is quite an interesting contrast
to the case when $\eta$ is the equality relation on $\mathbb N$, for
which vacillatory learning is known to be strictly more powerful than
explanatory learning \cite[Theorem 3.3]{Cas99}.  Furthermore, we
extend the ``non-union property'' of explanatory and behaviourally
correct learning (that is, the family of explanatorily
(resp.~behaviourally correctly) learnable classes is not closed under
union) \cite{Go67} by showing that for any positive equivalence
relation $\eta$, explanatory learning is not closed under union with
respect to $\eta$-families; we also obtain a characterisation of all
positive equivalence relations $\eta$ for which behaviourally correct
learning is closed under union with respect to $\eta$-families (see
Theorem \ref{thm:nonunionex}).  The current work thus serves a dual
purpose: first, casting inductive inference results in a more general
framework, where the elements of the domain over which concepts are
defined can be interrelated via any positive equivalence relation;
second, modelling the learning of r.e.\ structures endowed with any
r.e.\ equality relation.   

\section{Preliminaries}\label{sec:not}

\noindent
Any unexplained recursion-theoretic notation may be found in
\cite{Od89,Rog67,Soa87}.  The set of natural numbers
is denoted as $\N =\{0,1,2,\ldots\}$;
the set of all \emph{partial recursive
functions} and of all \emph{recursive functions} of one, and two
arguments over $\N$ is denoted by $\cP$, $\cP^2$, $\cR$ and $\cR^2$
respectively.  Any function $\psi \in \cP^2$ is called a
\emph{numbering of partial-recursive
functions} --- this numbering may or may not include all partial
recursive functions. Moreover, let 
$\psi \in \cP^2$; the notion $\psi_e$ stands for $\lambda x.\psi
(e,x)$ and $\cP_{\psi} = \{\psi_e \mid e\in \N\}$.
A numbering $\f \in \cP^2$ is
said to be an \emph{acceptable numbering} or \emph{G\"odel
numbering} of all partial recursive functions if $\cP_{\f} = \cP$ and
for every numbering~$\psi \in \cP^2$, there is a $c\in \cR$ such that
$\psi_e = \f_{c(e)}$ for all $e\in\N$ (see \cite{Rog67}).  Throughout
this paper, $\varphi_0,\varphi_1, \varphi_2,\ldots$ is a fixed
acceptable numbering of all partial recursive functions and
$W_0,W_1,W_2,\ldots$ is a fixed \emph{numbering of all recursively 
enumerable sets} (abbr.~r.e.~sets) of natural numbers, where $W_e$ is
the domain of $\varphi_e$ for all $e\in\N$. 

Let $e,x\in\N$; if $\varphi_e(x)$ is defined then we say that
$\varphi_e(x)$ \emph{converges}.  Otherwise, 
$\varphi_e(x)$ is said to \emph{diverge}.  Furthermore, if the
computation of $\varphi_e(x)$ halts within $s$ steps of computation
then we write $\varphi_{e,s}(x)\conv=\varphi_e(x)$; otherwise
$\varphi_{e,s}(x)$ diverges.  For all $e,s\in\N$ the set $W_{e,s}$ is
defined as the domain of $\varphi_{e,s}$.  Given any set $S$, 
$S^*$ denotes the set of all finite sequences of elements from~$S$. 
By $D_0,D_1,D_2,\ldots$ we denote any fixed canonical indexing of all
finite sets of natural numbers.  \emph{Turing reducibility} is denoted
by $\leq_T$; $A\leq_T B$, where $A, B\subseteq\N$, holds if $A$ can be
computed via a Turing machine which uses $B$
as an oracle; that is, it can give information on whether or not $x$
belongs to $B$.  $A\equiv_T B$ means that $A\leq_T B$ and $B\leq_T A$
both hold.  The symbol $\K$ denotes the \emph{diagonal halting problem}, i.e.,  
$\K=\{e: e\in\N,\; \varphi_e(e)\conv\}$.  Moreover, the \emph{jump} of
$K$, denoted by $K'$, is the relativised halting problem
$\{e:e\in\N,\; \varphi_e^K(e)\conv\}$, where
$\varphi_0^K,\varphi_1^K,\varphi_2^K,\ldots$ is an acceptable
numbering of all partial-$K$-recursive functions, which are partial
functions depending on the input as well as on answers to queries of
the form ``is $x$ in $K$?''.

For $\sigma\in (\N\cup \{\#\})^{\ast}$ and $n\in\N$ we write
$\sigma(n)$ to denote the element in the $n$th position of $\sigma$.
For any finite sequence $\sigma$ we use $|\sigma|$ to denote the
length of $\sigma$.  
Further, whenever $n \leq |\sigma|$, $\sigma[n]$ denotes the sequence
$\sigma(0),\sigma(1),\ldots,\sigma(n-1)$.  The concatenation of two
sequences $\sigma$ and $\tau$ is denoted by $\sigma\circ\tau$; for
convenience, and whenever there is no possibility of confusion, this
is occasionally denoted by $\sigma\tau$. 

A class $\cC$ is said to be \emph{uniformly r.e.} (or just
\emph{r.e.}) if there is an r.e.\ set 
$S\subseteq \N$ such that $\cC =\{W_i : i\in S\}$.  A class is said to
be \emph{one-one r.e.}, if the r.e. set $S$ as above additionally
satisfies the condition that for $i,j \in S$, $W_i=W_j$ iff $i=j$.  An
r.e. class 
$\cC=\{B_0,B_1,\ldots\}$ is said to be \emph{uniformly recursive} or
an \emph{indexed family} if there exists a recursive function~$f\in
\cR^2$ such that for all $i,x \in \N$, if $x \in B_i$ then $f(i,x)=1$
else $f(i,x)=0$.

\section{Learnability}\label{sec:learn}

\noindent
Recursion-theoretic notation mainly follows the textbooks of Odifreddi
\cite{Od89}, Rogers \cite{Rog67} and Soare \cite{Soa87}.  Background
on inductive inference may be found in \cite{JORS99}.  Let $\cC$ be a
class of r.e. languages.  Throughout this paper, the mode of data
presentation is that of a \emph{text}. A text is any 
infinite sequence of natural numbers and the $\#$ symbol, where the
symbol $\#$ indicates a pause in the data presentation. More formally,
a \emph{text} $T_L$ for a language $L \in \cC$ is any total mapping 
$T_L:\N \to \N \cup \{\#\}$ such that $L = \rng(T_L)\sm \{\#\}$.  We
use $\cnt(T)$ to denote the set $\rng(T)\sm \{\#\}$, i.e., the content
of a text $T$ contains only the natural numbers appearing in~$T$.
Furthermore, for every $n\in\N$ we use $T[n]$ to denote the finite
sequence $T(0),\ldots,T(n-1)$, i.e., the \emph{initial segment} of
length $n$ of $T$.  Analogously, for a finite sequence $\sigma \in
(\N\cup \{\#\})^{\ast}$ we use $\cnt(\sigma)$ to denote the set of all
numbers in the range of $\sigma$.

\begin{desc} \label{de:general} \rm
Further basic ingredients of the notions considered in this paper
are as follows.
\begin{enumerate}[(1)]
\item For each positive equivalence relation $\eta$, one can define
      an infinite sequence $a_0,a_1,\ldots$ of least representatives of the
      equivalence classes where each $a_n$ is the ascending limit of
      approximations $a_{n,t}$ where $a_{n,t}$ is the least natural number
      which is not $\eta_t$-equivalent to any $a_{m,t}$ with $m < n$
      (where $\eta_t$
      denotes the approximation to $\eta$ after $t$ steps of enumeration
      and is closed under reflexivity and transitivity).
      Alternatively, one can obtain $\eta$ from a construction of such
      a sequence where the $a_{n,t}$ approximate the $a_n$ from below
      and whenever an $a_{n,t+1} \neq a_{n,t}$ then $a_{n,t+1} = a_{m,t}$
      for some $m > n$ and whenever $a_{n,t}$ is not in the list at $t+1$
      then it is put into the equivalence class of some $a_{m,t}$ with $m<n$.
      Some algorithms to construct the equivalence relation $\eta$ explain
      on how to update these approximations to $a_0,a_1,\ldots$ and one
      should note that (the construction implies) the limit satisfies
      $a_0 < a_1 < \ldots$ and
      that for each $n$ there are only finitely many $t$ with $a_{n,t}
      < a_{n,t+1}$.

\item The classes whose learnability are considered are given by
      a uniformly r.e.\ one-one numbering of sets $B_0,B_1,\ldots$
      where each set $B_k$ is a union of $\eta$-equivalence classes;
      however, the indices $k$ of $B_k$ are usual natural numbers
      and not equivalence-classes of $\eta$. Such a family is called
      an {\em $\eta$-family} below and note that $\eta$-families
      are always infinite.

\item Infinite indexed families as
      considered by Angluin 
      \cite{An80} are too restrictive, as they might
      not exist for some $\eta$; however, every infinite indexed family
      has a one-one numbering and thus using the notion of infinite
      uniformly r.e.\ one-one numberings is the adequate choice for
      the present work.

\item The learner sees an infinite sequence $x_0,x_1,\ldots$ of members
      of one set $B_k$ (such sequences are called texts and can have
      pauses represented by a special pause symbol $\#$) and the learner
      has to find in the limit an r.e.\ index for $B_k$, which may or
      may not be equal to $k$. 

\item The hypotheses issued by the learners
      are always indices from a hypothesis space given as
      a numbering of $\eta$-closed r.e.\ sets; the most general hypothesis
      space is that of the $\eta$-closures of the members of a given
      acceptable numbering of all r.e.\ sets.

\item The present work focusses on the following learning criteria
      \cite{Bar74,Cas99,CL82,Fe72,Go67,JORS99,Od99,OSW86}:
      {\em Explanatory learning}, where the
       hypotheses of the learner converge on every text for a set $B_k$
      to a single index of $B_k$; 
      {\em Confident learning}, which is
      explanatory learning with the additional requirement that
      the learner also on texts not belonging to any language in the class
      has to converge to some index; 
      {\em Behaviourally correct learning}, 
      which is more general than explanatory learning and where the learner
      is only required to output on any text for $B_k$ almost always
      an index for $B_k$ but these indices can all be different;
      {\em Vacillatory learning}, where a learner is vacillatory
      iff it is a behaviourally correct learner for the class, with the
      additional constraint that on every text for a language $B_k$ in the
      class, the set of all hypotheses issued in response to this text is
      finite (thus, some of these hypotheses are output infinitely 
      often).
\end{enumerate}
\end{desc}

\noindent 
We now provide formal definitions of these criteria as well as the
criterion of \emph{finite} learning (sometimes known as \emph{one-shot
learning} in the literature; see \cite{Go67,TB75}), which is a more
restrictive version of explanatory learning.  In the following
definitions, a learner $M$ is a recursive function mapping
$(\N\cup\{\#\})^*$ into~$\N\cup\{?\}$; the $?$ symbol allows $M$ to
abstain from conjecturing at a stage.
If $M$ is presented with a text $T$ for any $\eta$-closed set $L$, it
is enough to assume that $\cnt(T)$ contains \emph{at least one}
element of each $\eta$-equivalence class contained in $L$; since
$\eta$ is r.e., $M$ on $T$ could simulate a complete text for $L$ by
enumerating at each stage $s$ the $s$-th approximation of the current
input's $\eta$-closure.      

The main learning criteria studied in this paper are \emph{explanatory
learning} (also called \emph{learning in the limit}) introduced by
Gold \cite{Go67} and \emph{behaviourally correct learning}, which goes back 
to Feldman \cite{Fe72}, who called it \emph{matching in the limit};
later, it was also studied by B\=arzdi\c n\u s, Case, Lynes and Smith
\cite{Bar74,Bar77,CL82,CS83}.  A fairly natural learning constraint on
these criteria is \emph{conservativeness} \cite{An80}, which requires
that any syntactic (resp.~semantic) mind change by the
explanatory (resp.~behaviourally correct) learner occur only if the
learner's original conjecture does not account for all the data
revealed in the text so far.

Furthermore, we will also consider \emph{finite learning} (see
\cite{Go67,TB75}), which is sometimes known as \emph{one-shot
learning} in the literature.  Other learning criteria studied in this
paper include \emph{confident learning} (see \cite{JORS99,OSW86}),
which requires the learner to converge syntactically on every
text for any language (even if it is outside the class to be learnt),
and \emph{vacillatory learning} (see \cite{Cas99}), according to which
a learner is permitted to vacillate between finitely many correct
indices in the limit.

\begin{defn}[Angluin \cite{An80}, %
B\=arzdi\c n\v s \cite{Bar74}, Case and Smith \cite{CS83}, %
Feldman \cite{Fe72}, Gold \cite{Go67}, Osherson, Stob and Weinstein %
\cite{OSW86}, Trakhtenbrot and B\=arzdi\c n\v s \cite{TB75}] \label{defnmain}
\rm Let $\cC$ be any class of r.e.\ languages.
\begin{enumerate}[(1)]
\item
$M$ \emph{explanatorily (\Ex) learns}
$\cC$ if, for every $L$ in $\cC$ and each text
$T_L$ for $L$, there is a number $n$
for which $L = W_{M(T_L[n])}$ and,
for every $j\geq n$, $M(T_L[j]) = M(T_L[n])$. 

\item
$M$ \emph{behaviourally correctly (\BC) 
learns} $\cC$ if, for every $L$ in $\cC$ 
and each text $T_L$ for $L$, there is a number $n$ for which 
$L = W_{M(T_L[j])}$ whenever $j\geq n$.

\item
$M$ \emph{finitely (\Fin) learns} $\cC$ if, 
for every $L$ in $\cC$ and each text
$T_L$ for $L$, there is a number $n$
for which $L= W_{M(T_L[n])}$ and 
for every $m< n$, $M(T_L[m]) =\, ?$ and for every~$j\geq n$,
$M(T_L[j]) = M(T_L[n])$.

\item
$M$ \emph{confidently (\Conf) learns}
$\cC$ if $M$ \Ex\ learns $\cC$ and $M$ converges on every text
for any language, that is, for every $L \subseteq \N$ and text $T_L$ for $L$,
there is a number $n$ such that for every $j \geq n$, $M(T_L[j]) =
M(T_L[n])$.   

\item $M$ \emph{vacillatorily (\Vac) learns} $\cC$ if $M$ \BC\ learns $\cC$ and
for every $L$ in $\cC$ and each text $T_L$ for $L$, $\{M(T_L[n]): n \geq 1\}$
is finite.

\item
$M$ \emph{conservatively explanatorily learns}
$\cC$ if $M$ \Ex\ learns $\cC$ and for all $n \in \N$, $k \geq 1$,
$M(T[n]) \neq M(T[n+k])$ only if $\cnt(T[n+k]) \not\subseteq W_{M(T[n])}$.

\item
$M$ \emph{conservatively behaviourally correctly learns}
$\cC$ if $M$ \BC\ learns $\cC$ and for all $n \in \N$, $k \geq 1$,
$W_{M(T[n])} \neq W_{M(T[n+k])}$ only if $\cnt(T$ $[n+k]) \not\subseteq 
W_{M(T[n])}$.
\end{enumerate}
\end{defn}

\noindent
We will also consider the learning constraint of \emph{monotonicity}
\cite{Jan91,LZ93b,Wie91}.  Furthermore, we give some results
concerning the dependence of learnability on the choice of hypothesis space.  
When considering learnability based on the choice of hypothesis space
$\{H_0,H_1,H_2,\ldots\}$, one replaces the hypotheses $W_0, W_1, \ldots$
in Definition~\ref{defnmain} by $H_0, H_1, \ldots$.
This work studies three types of requirements on the learner's
hypothesis space: the hypothesis space can be {\em class-comprising},
{\em class-preserving} or \emph{exact} \cite{LZ93a}.

A learner $M$ is said to be \emph{monotonic} \cite{Wie91} if, for all
texts $T$ for some language in the target class, whenever $M$ on input
$T[n]$ outputs $j_n$ and then at some subsequent step outputs
$j_{n+k}$ on input $T[n+k]$, the condition $W_{j_n} \cap \cnt(T)
\subseteq W_{j_{n+k}} \cap \cnt(T)$ holds. 

A learner $M$ is said to be \emph{strongly monotonic} \cite{Jan91} if,
for all texts $T$, whenever $M$ on input $T[n]$ outputs $j_n$ and then
at some subsequent step outputs $j_{n+k}$ on input $T[n+k]$, 
the condition $W_{j_n} \subseteq W_{j_{n+k}}$ holds.

A learner $M$ is said to be \emph{weakly monotonic} \cite{LZ93b} if,
for all texts $T$, whenever $M$ on input $T[n]$ outputs $j_n$ and then
at some subsequent step outputs $j_{n+k}$ on input $T[n+k]$, the
condition $\cnt(T[j+k]) \subseteq W_{j_n} \Rightarrow W_{j_n}
\subseteq W_{j_{n+k}}$ holds.  

A learner $M$ is said to be \emph{class-comprising} \cite{LZ93a,LZ93c}
if it learns the target class 
$\cC$ with respect to at least one hypothesis space
$\{H_0,H_1,\ldots\}$; note that learnability automatically implies
$\cC \subseteq \{H_0,H_1,\ldots\}$.

Unless otherwise stated, the learner's hypothesis space will be taken
as the $\eta$-closure of $W_0,W_1,\ldots$; in other words, the learner is
class-comprising.  For the notions of class-preserving and exact
learning, one considers some other recursively enumerable hypothesis space
$H_0,H_1,\ldots$ for interpreting the hypothesis of the learner. Only
uniformly recursively enumerable hypothesis spaces are considered.

A learner $M$ is said to be \emph{class-preserving} \cite{LZ93a,LZ93c}
if it learns the target class 
$\cC$ with respect to at least one hypothesis space
$\{H_0,H_1,\ldots\}$ such that $\{H_0,H_1,\ldots\} = \cC$.

A learner $M$ is said to be \emph{exact} \cite{LZ93a,LZ93c} if it
learns the target class 
$\cC := \{L_0,L_1,\ldots\}$ using $\{L_0,L_1,\ldots\}$ as its
hypothesis space (in other words, for all 
$i \in \N$, the learner outputs $i$ to signal that its conjecture is $L_i$).

\medskip
\noindent
Lange and Zeugmann \cite{LZ93c} showed that for learning uniformly
recursive families, learnability is to some extent independent of the
choice of hypothesis space in that exact, class-preserving and
class-comprising explanatory learnability all coincide.  Jain, Stephan
and Ye \cite{JSY09} observed more generally that for any finite
(resp.~explanatorily) learnable uniformly r.e.\ family $\cC$ and
hypothesis space 
$\mathcal{H} := \{H_0,H_1,\ldots\}$ such that $\mathcal{H}$ and $\cC$
consist of the same set of members (possibly ordered differently), one
can uniformly construct a finite (resp.~explanatory) learner for 
$\cC$ with respect to $\cH$ from an r.e.\ index of $\{\spn{d,x}: x \in
H_d\}$.  Thus, as the present paper deals exclusively with uniformly
r.e.\ families, it will be assumed throughout that finite
(resp.~explanatory) learning, class-preserving finite
(resp.~explanatory) learning and exact finite (resp.~explanatory)
learning are all equivalent (this equivalence may, however, fail if
other learning constraints are imposed at the same time).    

In some cases we consider learners using oracles.
In this case the learning criterion $I$ when the learners
are allowed use of oracle $A$ is denoted by $I[A]$.

For every learning criterion $I$ considered in the present paper, 
there exists a recursive enumeration~$M_0, M_1, \ldots$ of
the learning machines such that if a class is $I$-learnable,
then some $M_i$ $I$-learns the class.
We fix one such enumeration of learning machines.
We will also consider combinations of various learning criteria;
for example, one could require a learner to be both confident
and behaviourally correct.

\medskip
\noindent
Throughout this work, we only consider positive equivalence relations that
induce \emph{infinitely} many equivalence classes.
For any positive equivalence relation $\eta$ and $x \in \N$,
let $[x]$ be $\{y: y\,\eta\,x\}$.
Furthermore, for any $D \subseteq \N$, $[D]$ denotes $\bigcup_{x \in D}
[x]$.  For any finite $\{i_0,\ldots,i_n\} \subseteq \N$, 
the set $[\{a_{i_0},a_{i_1},\ldots,a_{i_n}\}]$ will simply be denoted by 
$[a_{i_0},a_{i_1},\ldots,$ $a_{i_n}]$.
An \emph{$\eta$-family} $\cC$ is a uniformly r.e.\ one-one infinite family, 
each of whose members is a union of $\eta$-equivalence classes.
Note that uniformly recursive infinite families might not exist for some $\eta$
and therefore an $\eta$-family is the nearest notion to a uniformly recursive
family which exists for each positive equivalence relation $\eta$.
A set is \emph{$\eta$-infinite} (resp.~$\eta$-finite) if it is equal 
to a union of infinitely (resp.~finitely) many $\eta$-equivalence classes;
note that an $\eta$-infinite set may not necessarily be recursively enumerable.
A set is \emph{$\eta$-closed} if it is either $\eta$-finite or $\eta$-infinite. 
In this paper, all families are assumed to consist of only $\eta$-closed
sets (for some given $\eta$).
For brevity's sake, we do not use any notation to indicate the
dependence of $a_n$
on $\eta$; the choice of $\eta$ will always be clear from the context.
A family $\mathcal{A}$ of sets is called a \emph{superfamily} of another
family $\mathcal{B}$ of sets iff $\mathcal{A} \supseteq \mathcal{B}$.

A useful notion that captures the idea of the learner converging on a
given text is that of a \emph{locking sequence}, or more generally that of a 
\emph{stabilising sequence}.
A sequence $\sigma \in (\N \cup \{\#\})^*$ is called a
\emph{stabilising sequence} \cite{Fulk85} for a learner $M$ on some language 
$L$ if $\cnt(\sigma) \subseteq L$ and for
all $\tau \in (L \cup \{\#\})^*$, $M(\sigma) = M(\sigma\circ\tau)$.
A sequence $\sigma \in (\N \cup \{\#\})^*$ is called a \emph{locking sequence}
\cite{BB75} for a learner $M$ on some language $L$ if $\sigma$ is a stabilising 
sequence for $M$ on $L$ and $W_{M(\sigma)} = L$.  The following proposition 
due to Blum and Blum \cite{BB75} will be occasionally useful.

\begin{prop}[Blum and Blum \cite{BB75}] \label{prop:lockingsequence}
If a learner $M$ explanatorily learns some language $L$, then there exists a
locking sequence for $M$ on $L$.  Furthermore, all stabilising sequences for
$M$ on $L$ are also locking sequences for $M$ on $L$.
\end{prop}

\noindent
The following theorem due to Kummer \cite{Ku90} will be useful for showing 
that a given family of r.e.\ sets has a one-one numbering.

\begin{thm}[Kummer \cite{Ku90}] \label{thm:kummeroneonenum}
Suppose $L_0,L_1,L_2,\ldots$ and $H_0,$ $H_1,H_2,\ldots$ are two
numberings such that the following conditions hold:
\begin{enumerate}[(1)]
\item for all $i,j \in \N$, $L_i \neq H_j$;
\item $H_0,H_1,H_2,\ldots$ is a one-one numbering;
\item for all $i \in \N$ and all finite $D \subseteq L_i$, there are infinitely
many $j$ such that $D \subseteq H_j$.
\end{enumerate}
Then $\{L_i: i \in \N \} \cup \{H_j: j \in \N \}$ has a one-one numbering.
\end{thm}

\section{Results for all Positive Equivalence Relations: \Fin, \Conf, \Ex, 
\Vac\ and \BC\ Learning}\label{sec:allposeq}

\noindent
In the present section, we investigate the relationship between the main 
learning criteria -- namely, finite, confident, explanatory, vacillatory
and behaviourally correct learning --  with respect to families closed under 
any given positive equivalence relation.  The first part of this section
will study, for any general positive equivalence relation $\eta$, the 
learnability of a particular $\eta$-family known as the \emph{ascending 
family for $\eta$}.  As will be seen later, the ascending family provides a 
useful basis for constructing $\eta$-families that witness the separation of 
various learnability notions. 

\begin{defn}\label{defn:ascendingfamily}
For all $n \in \N$, $A_n$ denotes the set 
$[a_0,a_1,...,$ $a_{n-1}]$.
The family $\{A_n: n \in \N\}$ will be denoted by 
$\cA_{\eta}$, and is called the \emph{ascending family for $\eta$}.  
\end{defn} 

\noindent
Note that each member of $\cA_{\eta}$ is $\eta$-finite; furthermore, 
$\cA_{\eta}$ is an $\eta$-family because $\eta$ induces infinitely many
equivalence classes and for all $n$, $a_n$ can be approximated from
below (c.f.~Description
\ref{de:general}, item (1)).  $\cA_{\eta}$ is analogous to the family
$\mbox{INIT} := \{\{y: y < x\}: x \in \mathbb N \}$ defined in
\cite[Section 4.3]{Jain96}. 
For brevity's sake, we do not use any notation to indicate the dependence of
$A_n$ on $\eta$; the choice of $\eta$ will always be clear from the context.

In the second part of this section, we study the question of whether the
learning hierarchy
$$
\Fin \subset \Conf \subset \Ex \subset \Vac \subset \BC.
$$
is strict for the class of $\eta$-families (for any given positive equivalence 
relation $\eta$).
It turns out that while the two chains of inclusions $\Fin \subset
\Conf \subset \Ex$
and $\Vac \subset \BC$ hold for \emph{all} positive equivalence relations,
there is a positive equivalence relation $\vartheta$ for which every
vacillatorily learnable $\vartheta$-family is also explanatorily learnable.
The construction of $\vartheta$ will be given in the next section.   
We begin with a few basic examples of $\eta$-families to illustrate
some of the notions introduced
so far. 

\begin{exmp}[Ershov \cite{Er74}]\label{exmp:recursiveset}
If $A$ is a recursive and coinfinite set, then $x\, \eta_A\, y\,
\Leftrightarrow \,( x=y \vee (x \in A$ $\wedge$ $y \in
A))$ is a positive equivalence relation.  $\cF := \{A \} \cup
\{\{x\}: x \notin A\}$ is an $\eta_A$-family
since (1) every equivalence class of $\eta_A$ is either $A$ or a
singleton $\{x\}$ with $x \notin A$,
which implies that $\cF$ is infinite and each member of $\cF$ is
$\eta_A$-closed, and (2) there is a uniformly recursive one-one
numbering $\{F_i\}_{i \in \N}$ of $\cF$; for example, one could set
$F_0 = A$ and $F_{i+1} = \{x_i\}$ for all $i$, where
$x_1,x_2,x_3,\ldots$ is a one-one recursive enumeration of $\N \sm A$.
 $\cF$ is also
finitely learnable via a learner that outputs ? until it sees the
first number $x$ in the input;
if $x \in A$ then $A$ is conjectured, and if $x \notin A$ then $\{x\}$
is conjectured.    
\end{exmp}

\begin{exmp}[Ershov \cite{Er74}]\label{exmp:symmetricdifference}
If $R$ is an r.e.\ set and $D_0,D_1,D_2,\ldots$ is a one-one numbering
of all finite sets, 
then $x\, \eta_R\, y \Leftrightarrow D_x \triangle D_y \subseteq R$ is
a positive
equivalence relation ($\triangle$ denotes the symmetric difference). 
If $S \cap R = \emptyset$,
then $L_S := \{x: D_x \cap S \neq \emptyset\}$ is $\eta_R$-closed. 
Suppose $\N \sm R$ contains an infinite
r.e.\ set $C$.  Let $\cF$ consist of all sets $L_{C'}$ such that $C' =
C \sm F$ for some finite set $F$.
Then $\cF$ is an $\eta_R$-family that is not behaviourally correctly
learnable.        
\end{exmp}

\noindent
The next theorem shows
that for any positive equivalence relation $\eta$, the ascending family
witnesses that explanatory learning is strictly more 
powerful than confident learning.    

\begin{thm}\label{thm:ascendexnotconf}
For every positive equivalence relation $\eta$,
the ascending family $\cA_{\eta}$ is 
explanatorily learnable but not confidently learnable.  
One can add the set $\mathbb N$ to $\cA_{\eta}$
and obtain an $\eta$-family which is not 
behaviourally correctly learnable.
\end{thm}

\def\proofofascendexnotconf{
\ownproof
Note that all members of the ascending family are $\eta$-finite and for each
member $A_n$, the learner can find in the limit $n$ as follows: At
each time $t$, the learner takes $n_t$ -- the $t$-th approximation
to $n$ -- to be the least number such that $a_{n_t,t}$ has not yet 
appeared in the text and this $n_t$ converges to $n$: From the time 
$t$ onwards where all $a_{m,t}$ with $m \leq n$ have converged to 
their final value $a_m$ and furthermore all $a_m$ with $m < n$
have appeared in the text, it holds that $n_t = n$ and that therefore the
conjecture is $A_n$ which can be given by a default index for this set. 

Now assume, by way of contradiction, that $\cA_{\eta}$ has a
confident learner $N$.  By \cite[Exercise 5--14]{JORS99}, there is some
$\sigma \in \N^*$ such that for all $\tau \in \N^*$, $N(\sigma\tau) = 
N(\sigma)$.  Since $A_0 \subset A_1 \subset A_2 \subset \ldots$ and
almost all of the $A_i$'s contain $\cnt(\sigma)$,
one can pick a least $n$ such that $W_{N(\sigma)} \neq A_n$
and $\cnt(\sigma) \subseteq A_n$.  Then $\sigma$ could be extended 
to a text $T$ for $A_n$ such that $N$ on $T$ converges to $N(\sigma)$ -- 
an incorrect index for $A_n$, contradicting the fact that $N$ confidently 
learns $\cA_{\eta}$.    
\sp
Furthermore, if one adds $\mathbb N$ to $\cA_{\eta}$,
the resulting family
contains $\mathbb N$ and a strictly ascending sequence of sets whose union
is $\mathbb N$, thus by Gold 
\cite{Go67} the family is not learnable from
positive data. This applies to all learning criteria considered in this
paper.~\qed
}
\proofofascendexnotconf

\medskip
\noindent
The following proposition provides a method for establishing that a given
uniformly r.e.\ family is an $\eta$-family. 

\begin{prop}\label{prop:ascendsuperfamily}
Every uniformly r.e.\ superfamily of $\cA_{\eta}$ that
consists of $\eta$-closed sets is an $\eta$-family;
in particular, the families of all $\eta$-finite sets and
all $\eta$-closed r.e.\ sets are $\eta$-families.
\end{prop}

\def\proofofascendsuperfamily{
\ownproof
According to Kummer's theorem (cf.~Theorem \ref{thm:kummeroneonenum}),
a union of two disjoint uniformly r.e.\ families
has a one-one numbering whenever every finite subset of any member of
one family has infinitely many supersets in the 
other family and the 
latter family has a one-one numbering. So let $E_0,E_1,\ldots$ be the
given r.e.\ 
family and let $A_0,A_1,\ldots$ be the default one-one enumeration of the 
ascending $\eta$-family. Now one chooses for the second family $\cF_2$
the family 
of all sets $A_{2k}$ and for the first family $\cF_1$ all those sets
$E_k$ which are 
not in $\cF_2$.
Note that all sets $A_{2k+1}$ are therefore in $\cF_1$. 
For completing the proof, one has to show that $\cF_1$ 
is a uniformly r.e.\ family.
\sp
Let $E_{k,t}$ denote the $t$-th approximation to $E_k$.
For this one defines sets $U_{k,n,t}$ as follows: 
\begin{quote}
If (i) $a_{n,t} \notin E_{k,t}$
and (ii) either $n$ is odd and all $a_{m,t} \in E_{k,t}$ with $m<n$ or
$a_{n+1,t} \in E_{k,t}$,

then one enumerates into $U_{k,n,t}$ all elements of
$E_{k,t}$ until either $a_{n,t}$ is enumerated into $E_{k,s}$ for some
$s \geq t$ or there is an $m \leq n+1$ and $s \geq t$
with $a_{m,s} \neq a_{m,t}$.

If this happens, one lets $h$ be the largest element enumerated into $U_{k,n,t}$
so far and one enumerates into $U_{k,n,t}$ all elements of $A_{2h+1}$ which
contains all elements enumerated into $U_{k,n,t}$ so far.
\end{quote}
Note that $A_{2h+1}$ mentioned at the end of this algorithm
is not a member of the second numbering $\cF_2$.
It is easy to see that the
numbering of all
$U_{k,n,t}$ is a family of uniformly r.e.\ sets and that each of these sets
is either equal to $E_k$ in the case that the condition associated to
$(n,k)$ witnesses this (either $n$ is odd and $a_n$ is the least non-element
of $E_k$ or $a_n \notin E_k$ while $a_{n+1} \in E_k$) or the set is explicitly
made to be an $A_{2h+1}$ which is not contained in the second
numbering $\cF_2$ but in
the numbering of the $E_k$. There is only one set which is not captured by this
numbering but might be in the numbering of the $E_k$:
This is the set $\mathbb N$.
If this set is also in the numbering, then one has to add it afterwards
explicitly to $\cF_1$.
In the case that one has a family which does
not respect $\eta$, one can also for this achieve the result --- although it
is not the intention of the current work to consider such families --- by
adding the members of a further numbering of sets $\tilde U_{k,x,y,t}$ which
are equal to $E_k$ in the case that $x \, \eta_t \, y$ and $x \in E_{k,t}$
and $y \notin E_k$; in the case that this is not satisfied, the set
$\tilde U_{k,x,y,t}$ is made equal to some set of the form $A_{2h+1}$
when the violation of the condition is discovered.
\sp
As the class of all $\eta$-finite sets and the class of all
$\eta$-closed r.e.\ sets
are uniformly r.e.\ superfamilies of the ascending $\eta$-family, these
classes are also $\eta$-families.~\qed
}
\proofofascendsuperfamily

\medskip
\noindent
A minor modification of the proof of Proposition \ref{prop:ascendsuperfamily} 
reveals a slightly more general result: for any positive equivalence relation 
$\eta$ and any strictly increasing recursive enumeration $e_0,e_1,e_2,\ldots$, 
every uniformly r.e.\ superfamily of $\{A_{e_i}: i \in \N\}$ is an
$\eta$-family.  
This variant of Proposition \ref{prop:ascendsuperfamily} will be occasionally 
useful for showing that a given uniformly r.e.\ class is an $\eta$-family.

\begin{prop}\label{prop:subascendsuperfamily}
Let $f$ be any strictly increasing recursive function.  Then, for any
given positive equivalence relation $\eta$, every uniformly r.e.\ superfamily 
of $\{A_{f(i)}: i \in \N\}$ consisting of $\eta$-closed sets is an
$\eta$-family. 
\end{prop}     

\noindent
The next result shows that for any given positive equivalence relation $\eta$, 
behaviourally correct learning is more powerful than explanatory learning 
with respect to the class of $\eta$-families. 

\begin{thm} \label{th:bcnotex}
For every positive equivalence relation $\eta$, there is an 
$\eta$-family which is behaviourally correctly learnable but
not explanatorily learnable.
\end{thm}

\def\proofofbcnotex{
\ownproof
The $\eta$-family of all $\eta$-finite sets is behaviourally correctly
learnable: When $D$ is the set of all data items the learner has seen
so far, then it conjectures the set $\{x: \exists y \in D\,[x\,\eta\,y]\}$.
It is straightforward to compute indices of these sets from lists of elements
in the set $D$. Next, we distinguish two cases: first, when the family of all
$\eta$-finite sets is explanatorily learnable; second, when this family is not
explanatorily learnable.  In the case that this family is not
explanatorily learnable,
the proof is already completed; so assume that it has an explanatory
learner~$M$.  We construct another $\eta$-family that is behaviourally correctly
learnable but not explanatorily learnable.
\sp
One considers
the class which contains for each $n$ the following sets:
\begin{enumerate}[1.]
\item $A_n = \{x: \exists k < n\,[x\,\eta\,a_k]\}$;
\item $F_n = \{x: \exists k \neq n\,[x \, \eta \, a_k]\}$;
\item If $W_n$ is finite then for all $m \leq |W_n|$ and $m > n$ the set
      $B_{n,m} = \{x: \exists k < m$ $[k \neq n$ $\wedge$ $x \, \eta \, a_k]\}$.
\end{enumerate}
This class is a 
superfamily of the ascending $\eta$-family
$A_0,A_1,$ $\ldots$ and so by Proposition
\ref{prop:ascendsuperfamily}, it is sufficient 
to show that it is uniformly r.e.\ for proving that it is an $\eta$-family.
The enumeration-procedure checks for each combination of certain
indexing parameters
(described in detail below) whether these witness that the
corresponding set is in the class;
when an error is discovered with respect to the parameters, the enumerated set
is overwritten by enumerating some member of the ascending family which
contains all the numbers enumerated so far (this is possible because any
finite set $D$ is contained in the ascending family $\{x: \exists k <
\max(D) + 1\,[x \, \eta \, a_k]\}$,
where $\max(\emptyset)$ is defined to be $1$).  
\sp
We now describe in detail the enumeration procedures for (i) $A_n$,
(ii) $F_n$ and (iii) $B_{n,m}$.  
\begin{enumerate}[i.]
\item $A_n$ can be enumerated from the parameter $n$ and no check is needed.

\item One can, in the limit, compute for every $n$ the value $a_n$ and a
stabilising sequence $\sigma_n$ of $M$ for the set 
$[a_n]$.
$F_n$ can be enumerated based on $a_n$ and a
stabilising sequence $\sigma_n$ of the explanatory learner $M$ for
$[a_n]$; specifically, one has that $F_n = \{x: \exists \tau \in ([a_n] \cup
[x])^*\,[M(\sigma_n \circ x \circ \tau) \neq M(\sigma_n)]\}$.
Thus, at time $t$, one can make a guess $e_{n,t}$ for the enumeration 
procedure of $F_n$ based on the value of $a_{n,t}$
and a guess $\tau_s$ for the actual string $\sigma_n$, where,
given a default enumeration $\tau_0,\tau_1,\tau_2,\ldots$ of
all finite sequences,
$s$ is the least index $s'$ such that
$\tau_{s'} \in ([a_{n,t}]_t \cup \{\#\})^*$, 
and $\tau_{s'}$ ``appears'' to be a stabilising sequence of $M$ for
$[a_{n,t}]$, in the sense that for all $\tau' \in ([a_{n,t}]_t \cup 
\{\#\})^*$ of length at most $t$, 
$M(\tau_{s'} \circ \tau') = M(\tau_{s'})$ (if no such $\tau_{s'}$
exists, then set $\tau_s = \lambda$); here $[a_{n,t}]_t$ refers to the 
set of elements enumerated into $[a_{n,t}]$ up to and including stage 
$t$, based on some fixed enumeration of $[a_{n,t}]$.  
One may then choose $e_{n,t}$ such that $W_{e_{n,t}} = \{x: 
\exists \tau \in ([a_{n,t}] \cup [x])^*\,
[M(\tau_s \circ x \circ \tau) \neq M(\tau_s)]\}$. 
The set $F_n$ can be
enumerated from a pair $(n,t)$ which uses the guess of the index $e_{n,t}$
for the enumeration procedure of $F_n$ at time $t$ and which enumerates
$W_{e_{n,t}}$ until an $s>t$ is found with $e_{n,s} \neq e_{n,t}$; if that
latter happens, then the enumeration is stopped and some member of the
ascending family is enumerated which contains all the numbers so far enumerated.
For sufficiently large $t$, the procedure will enumerate from $(n,t)$ the
set $F_n$. 

\item For the set $B_{n,m}$, one considers all triples $(n,m,t)$ where
$W_{n,t}$ has at least $m$ elements and then one will enumerate for this
triple all elements equivalent to some
$a_{k,t}$ with $k < m  \wedge k \neq n$ until an $s > t$ is found
such that $a_{k,t} \neq a_{k,s}$ for some $k \leq m$ or
$W_{n,s} \neq W_{n,t}$. If that happens, the parameters are invalid
and the set is made equal to some member of the ascending family which
contains all the data enumerated so far.
\end{enumerate}

\noindent
The so defined family is a superset of all $\eta$-cosingletons $F_n$.
If the family was explanatorily learnable, then every $F_n$ would
have a locking sequence $\tau_n$ which can be found in the limit. If $W_n$
is finite, then $\tau_n$ must contain an element $h \geq |W_n|$,
due to $A_n$ being in the class, and 
$B_{n,|W_n|}$ being in the class and containing all elements
of $F_n$ strictly below $|W_n|$ and perhaps some others more.
Therefore one can check using the halting problem oracle $K$
whether $W_n$ enumerates at least $\max(range(\tau_n))+1$ elements.
If the answer is ``yes'' then $W_n$ must be infinite; if the answer is ``no''
then $W_n$ must be finite. Thus one could decide relative to $K$ whether $n$ is
in the set $\mathrm{FIN} = \{n : W_n$ is finite$\}$. But this is impossible, as
$\mathrm{FIN}$ is $\Sigma_2$-complete, while $K$ is only
$\Sigma_1$-complete---see
\cite[Proposition X.9.6]{Od89}.
Thus the family cannot be explanatorily learnable.  

It remains to show that the so constructed $\eta$-family
\pagebreak[3]
is behaviourally correctly learnable. So a behaviourally correct learner
\pagebreak[3]
$N$ would first find in the limit the following pieces of information,
\pagebreak[3]
the indices issued on the way to find these pieces of information
enumerate something, but might
not enumerate any 
set in the constructed $\eta$-family:
\begin{enumerate}[1.]
\item The least number $n$ such that $a_n$ does not appear in the text;
\item An index $e_n$ such that $W_{e_n} = F_n$.
\end{enumerate}
Furthermore, let $t$ be the number of input items (including repetitions
and pause symbols) from the text processed so far and $m = \min\{k > n:
a_{k,t}$ has not yet appeared in the text$\}$.
\sp
In the case that $m = n+1$, the learner conjectures
$A_n$. In the case that $m > n+1$, the learner checks whether
$W_{n,t}$ has more than $m$ elements.
\sp
If the answer is ``yes'', then the
learner conjectures an index for an r.e.\ set which first
starts enumerating all elements of $B_{n,m}$ until it happens that
there is an $s > t$ with $W_{n,s} \neq W_{n,t}$ or
$\exists k \leq m\,[a_{k,s} \neq a_{k,t}]$; if that happens
then the learner enumerates $W_{e_n}$ which is equal to $F_n$.
\sp
If the answer is ``no'', then the learner directly enumerates all
members of~$W_{e_n}$.
\sp
There is always one $n$ for
which $a_n$ is not contained in the set; it is clear that the learner finds
the least such $n$ in the limit; furthermore, in the case that it exists,
the learner also finds the second least $m$ such that $a_m$ is not in the
set in the limit. Now one of the following cases holds:
\begin{enumerate}[1.]
\item $m = n+1$: In this case, the learner will conjecture $A_n$ almost always,
thus it behaviourally correctly learns that set and this is the only set for
which $a_n$ and $a_m$ with the above condition are the two least non-elements.
\item $n+1 < m$ and $m \leq |W_n|$ and $W_n$ is finite:
In this case, for all sufficiently large $t$, $W_{n,t} = W_n$ and
$a_{k,t} = a_k$ for all $k \leq m$ and all $a_k$ with
$k < m \wedge k \neq n$ have appeared in the text; in these
circumstances, the learner will conjecture $B_{n,m}$ and this
set is correct.
\item $\max(n+1,|W_n|) < m$: This case does not occur, there is no set
with the two least non-elements $a_n$ and $a_m$ in the $\eta$-family
considered.
\item $m$ does not exist: Then the set to be learnt is $F_n$. When
$t$ is sufficiently large then it holds that either $W_n$ is infinite
and all $k \leq n+1$ satisfy the below statement $(*)$
or $W_n$ is finite and all $k \leq |W_n|$ satisfy $(*)$.
Here $(*)$ is the statement that $a_{k,t} = a_k$
and whenever $k \neq n$ then $a_{k,t}$ has appeared among the first $t$
input data items of the text. One can see that in both the ``either-case''
and the ``or-case'', the set conjectured by the learner is $F_n$.
\end{enumerate}
This case-distinction completes the proof by verifying that
$N$ is a correct behaviourally correct learner for the family.~\qed
}
\proofofbcnotex

\medskip
\noindent
Vacillatory learning, according to which a learner is allowed to switch 
between any finite number of correct indices in the limit, is known to 
be strictly weaker than behaviourally correct learning for general 
families of r.e.\ sets \cite{Cas99}.  The next main result -- Theorem
\ref{cor:bcnotvac} -- asserts that 
for any given positive equivalence relation $\eta$, this relation between 
the two criteria holds even for certain $\eta$-families.  We begin with 
the following proposition, from which the separation result may be 
deduced.  

\begin{prop} \label{pr:vacfinite}
If the class of $\eta$-finite sets is vacillatorily learnable then one
can relative to the halting problem $K$ compute a sequence $e_0,e_1,\ldots$
of characteristic indices of $\eta$-finite and $\eta$-closed sets
$E_0,E_1,\ldots$ which form a partition of $\mathbb N$.
\end{prop}

\def\proofofvacfinite{
\ownproof
Let $M$ be the vacillatory learner for the class of all $\eta$-finite sets.
Assume that all $E_m$ and $e_m$ with $m<n$ are computed. Now one finds with
help of $K$ the first $k$ such that $a_k \notin \bigcup_{m<n} E_m$. Furthermore,
there is a finite sequence $\sigma$ of elements $\eta$-equivalent to $a_k$ such
that any further finite sequence $\tau$ of elements $\eta$-equivalent to $a_k$
satisfies that $M(\sigma\tau)$ equals to $M(\rho)$ for some prefix $\rho$ of
$\sigma$; if such a sequence would not exist, one could construct a text
for the equivalence class of $a_k$ on which $M$ outputs infinitely many
different hypotheses. $M$ has issued at most $|\sigma|+1$ many different
indices on prefixes of $\sigma$. 
Now let $F_n$ be the union of the sets
$\bigcup_{m<n} E_m$ and the set of all $x$ for which there is
a $\tau$ consisting of elements which are $\eta$-equivalent to $a_k$ or to $x$
such that $M(\sigma x \tau)$ contains an index not in the finite set
$\{M(\rho): \rho \preceq \sigma\}$. Note that $F_n$ does not contain
any element $\eta$-equivalent to $a_k$ by the choice of $\sigma$;
furthermore, whenever for an $x$ there is no prefix $\rho \preceq \sigma$
with $W_{M(\rho)}$ enumerating $\{y: y \,\eta\, x$ $\vee$ $y \, \eta \, a_k\}$
then $x \in F_n$. Furthermore, it follows from the definition of $F_n$
that $F_n$ is $\eta$-closed. Now let
$R = \{\rho \preceq \sigma: W_{M(\rho)} \cap F_n = \emptyset\}$ and note that
$R$ can be determined using the halting problem $K$ as an oracle.
Note that $R$ is finite. The set $E_n = \bigcup_{\rho \in R} W_{M(\rho)}$
is $\eta$-finite and the complement of $F_n$; thus one can from the
enumeration-procedures of $F_n$ and $E_n$ compute a characteristic
index $e_n$ of $E_n$. Furthermore, the cardinality $c_n$ of $R$ is an
upper bound on the number of $a_k$ in $E_n$ and $E_n$ is $\eta$-finite.
As every $a_k$ appears in exactly one $E_n$ and
as all $E_n$ are $\eta$-closed, the $E_n$ form a partition of $\mathbb N$
into $\eta$-finite sets.~\qed
}
\proofofvacfinite

\def\proofofbcnotvac{
\ownproof
One can
use Proposition~\ref{pr:vacfinite} to construct, in the case
that the class of $\eta$-finite sets is vacillatorily learnable, the following
class which is behaviourally correctly learnable but not vacillatorily
learnable: The family contains
(i) all members $A_n$ of the ascending family;
(ii) all sets $F_n$ as constructed in the proof of 
Proposition~\ref{pr:vacfinite};
(iii) in the case that $W_n$ is finite, all sets $A_m \cap F_n$ with
      $m \leq |W_n|$.
For this class, one can show as in Theorem~\ref{th:bcnotex} that it is
a $\eta$-family and that it is behaviourally correctly learnable and
that it is not explanatorily learnable; the latter proof can be adjusted
to a proof that it is not vacillatorily learnable. To see this,
consider a vacillatory learner; for this learner one can compute using $K$
a sequence $\sigma$ of elements in $F_n$ such that every extension
$\sigma\tau$ of it using for $\tau$ only elements from $F_n$ satisfies
that $M(\sigma\tau) = M(\rho)$ for a prefix $\rho$ of $\sigma$. Now
if $W_n$ is finite then there are at least $|W_n|-c_n$ many sets of the
form $A_m \cap F_n$, $m \in \N$,
in the class (where $c_n$ is computed using $K$
as in Proposition~\ref{pr:vacfinite}) and for each of them, either some
$M(\rho)$ with $\rho \preceq \sigma$ must be an index for it or
$\sigma$ contains an element outside the set $A_m \cap F_n$. Note that
the smallest number outside $A_m$ is at least $m$. Thus if $W_n$ is finite
and if one takes $d_n = \max(range(\sigma))+|\sigma|+c_n$,
then $d_n \geq |W_n|$.
Therefore one can compute from $n$ the number $d_n$ using $K$ and then
check using $K$ whether $|W_n| \leq d_n$. If this is true then $W_n$
is finite else $W_n$ is infinite. Thus the existence of a vacillatory
learner would lead to an incorrect result about the arithmetic
hierarchy \cite[Proposition X.9.6]{Od89}
and therefore the $\eta$-family constructed is not vacillatorily learnable.~\qed
}

\begin{thm}\label{cor:bcnotvac}
For every positive equivalence relation $\eta$, there is an $\eta$-family which
is behaviourally correctly learnable but not vacillatorily learnable.
\end{thm}

\proofofbcnotvac

\medskip
\noindent
Moving down the learning hierarchy given at the start of the present 
section, the following theorem shows that for any positive equivalence 
relation $\eta$, finite learning can be more restrictive than confident 
learning with respect to $\eta$-families.

\begin{thm}\label{thm:finconfsepallper}
Let $\eta$ be any given positive equivalence relation such that there
is at least one finitely learnable $\eta$-family.  Then there is an
$\eta$-family that is confidently learnable but not finitely learnable.
\end{thm}

\def\proofoffinconfsepallper{
\ownproof
Let $B_0,B_1,\ldots$ be any finitely learnable $\eta$-family.
Now, for every set $B_n$ there is a finite subset
$C_n$ such that $B_n$ is the only superset in this family. These sets
$C_n$ must exist, as they are the data-items observed so far when the finite
learner conjectures $B_n$ from the default text of $B_n$ and as the conjecture
cannot be revised, there cannot be an $m \neq n$ with $C_n \subseteq B_m$.
Now one considers the family in which one replaces $B_2$ by $\mathbb N$.
This family has a confident learner which first issues the same conjectures
as the finite learner until it has seen all data in $C_0 \cup C_1$.
If that happens, $\mathbb N$ is the only consistent hypothesis and the
learner makes a mind change to that set. As this new class is not
inclusion-free, it is not finitely learnable.~\qed
}
\proofoffinconfsepallper

\section{The Non-Union Theorem and Finitely Learnable Classes}

\noindent
As Gold \cite{Go67} observed, the class consisting of $\N$ and all finite 
sets is not learnable in any sense considered in the present paper. Thus
adding only one set to the learnable class of all finite set makes it
unlearnable. This is not true for all classes. For example, in the case
that one considers languages made of strings, Angluin \cite[Example 1]{An80}
provided the class of all non-erasing pattern languages which can
be learnt from positive data explanatorily. If one adds a single set,
then the class is still explanatorily learnable, the reason 
being that the set of shortest strings in a non-erasing pattern language 
defines this pattern language uniquely; if now the set of shortest strings of 
the new language coincides with that of one of the members of the class, then 
one can add in one more string on which the two languages differ and use these
finitely many strings to detect the difference.

Similarly 
the union of two disjoint infinite explanatorily
learnable classes of r.e.\ languages may not even be behaviourally correctly 
learnable. Blum and Blum \cite{BB75} noted that the family of explanatorily 
(resp.~behaviourally correctly) learnable classes of recursive functions is 
also not closed under union. This section deals with the
question of whether the non-union property of explanatory (resp.~vacillatory, 
behaviourally correct) learning holds for the class of $\eta$-families, where 
$\eta$ is any given positive equivalence relation.  For any learning criterion 
$I$ and any positive equivalence relation $\eta$, say that $I$ is \emph{closed 
under union with respect to $\eta$} iff for any $\eta$-families $\cL$ and $\cH$ 
such that $\cL \cup \cH$ is an $\eta$-family, if $\cL$ and $\cH$ are
$I$-learnable, 
then $\cL \cup \cH$ is $I$-learnable.  Somewhat interestingly, while
explanatory and 
vacillatory learnability are not closed under union with respect to
any $\eta$, the 
answer for behaviourally correct learning depends on whether or not
there are at least two $\eta$-infinite r.e.\ sets.  

\begin{prop}\label{prop:natetainf}
Let $\eta$ be any given positive equivalence relation.  If $\N$ is the
only $\eta$-infinite r.e.\ set, then $\N$ is not contained in any behaviourally
correctly learnable $\eta$-family.
\end{prop}

\def\proofofnatetainf{
\ownproof
Assume that $\{B_0,B_1,...\}$ is a behaviourally correctly learnable
$\eta$-family 
containing $\N$.
By \cite[Corollary 3]{BCJ99}, there is a tell-tale set $H$ for
$\N$. Here a set
$S$ is said to be a {\em tell-tale} with respect
to $\cL$ for a language $L \in \cL$ if
$S \subseteq L$, $S$ is finite and
for all $L' \in \cL$, $S \subseteq L' \subseteq L$ implies
$L=L'$ (cf.~\cite{An80}).
Now there is a maximal subset $H'$ of $H$ such that $H'$ is contained in 
infinitely many $B_i$.  If now $H' = H$ then there is a $B_i \neq \N$
which contains $H$ and thus $H$ is not a tell-tale for $\N$.  If there
is an $x \in 
H \sm H'$ then there is an $n$ such that no $B_m$ with $m > n$ contains 
$H' \cup \{x\}$ as a subset.  Now one can one-one enumerate the union of all 
$B_m$ with $m > n$ such that $H' \subseteq B_m$; this union is an
$\eta$-infinite r.e.\ set which does not contain $x$,
but such a set does not exist. So this case does not hold either.~\qed
}
\proofofnatetainf

\begin{thm}\label{thm:nonunionex}
Let $\eta$ be any given positive equivalence relation. Then the following
hold.
\begin{enumerate}[(a)]
\item There are disjoint, explanatorily (resp.~vacillatorily) learnable 
$\eta$-families $\cL_1$ and $\cL_2$ for which $\cL_1 \cup \cL_2$ is an 
$\eta$-family that is not explanatorily (resp.~vacillatorily) learnable.  
\item Behaviourally correct learning is closed under union with respect
to $\eta$
iff $\N$ is the only $\eta$-infinite r.e.\ set.
\end{enumerate}      
\end{thm}

\def\proofofnonunionex{
\ownproof
We split the proof into two main cases.

Case 1:
$\N$ is the only $\eta$-infinite r.e.\ set.
Let $\cL_1$ and $\cL_2$ be any two behaviourally correctly learnable 
$\eta$-families such that $\cL_1 \cup \cL_2$ is an $\eta$-family.  By
Proposition 
\ref{prop:natetainf}, both $\cL_1$ and $\cL_2$ contain only
$\eta$-finite sets.  
Thus $\cL_1 \cup \cL_2$ contains only $\eta$-finite sets.  Note that
any class of 
$\eta$-finite sets is behaviourally correctly learnable: on input
$\sigma$, a \BC\ 
learner conjectures $\bigcup_{x \in \cnt(\sigma)}[x]$.  Therefore
$\cL_1 \cup \cL_2$ 
is also behaviourally correctly learnable, and so behaviourally correct 
learning is closed under union with respect to $\eta$.  This proves the
``if'' direction of statement (b).

\medskip
\noindent
Next, it is shown that explanatory (resp.~vacillatory) learning is not closed 
under union with respect to $\eta$.  
Set 
\begin{equation*}
\begin{split}
\cL_1 &= \{A_{2n+1}: n \in \N\} \cup \{[a_k]: k \in \N\}, \\
\cL_2 &= \{A_{2n+2}: n \in \N\} \cup \{[a_k,a_l]: k,l \in \N \wedge k < l\}.
\end{split}
\end{equation*}
Note that $\cL_1$ has the following uniformly r.e.\ numbering:
for all $n,s \in \N$,
\ifx\versy\fully
\begin{equation*}
\begin{split}
L_{2n} &= A_{2n+1}, \\
L_{2\spn{n,s}+1} &= [a_{n,s}].
\end{split}
\end{equation*}
\fi
\ifx\versy\confy
$ L_{2n} = A_{2n+1} \mbox{ and }$ $L_{2\spn{n,s}+1} = [a_{n,s}].
$
\fi
Similarly, $\cL_2$ has the following uniformly r.e.\ numbering 
$\{H_0,H_1,H_2,\ldots\}$.  For all $n \in \N$, set $H_{2n} =
A_{2n+2}$.  For all 
$m,n,s \in \N$, if $a_{m,s} = a_m$ and $a_{m+n+1,s} = a_{m+n+1}$, set 
$H_{2\spn{m,n,s}+1} = [a_{m,s},a_{m+n+1,s}]$, and if there is a least $t > s$ 
such that $a_{m,t} \neq a_{m,s}$ or $a_{m+n+1,t} \neq a_{m+n+1,s}$, set 
$H_{2\spn{m,n,s}+1} = A_{2n'+2}$, where $n'$ is the least $k$ such that 
$A_{2k+2}$ contains all the elements enumerated into $H_{2\spn{m,n,s}+1}$ up 
until stage $t$. A similar proof shows that $\cL_1 \cup \cL_2$ has a uniformly 
r.e.\ numbering. Thus by Proposition \ref{prop:subascendsuperfamily},
$\cL_1,\cL_2$ and $\cL_1 \cup \cL_2$ are $\eta$-families.

\medskip 
\noindent
The following learner $M$ explanatorily learns $\cL_1$.  On input $\sigma$, 
$M$ first checks whether or not $a_0 \in \cnt(\sigma)$.  If $a_0 \in 
\cnt(\sigma)$, then $M$ conjectures $A_{2n+1}$ for the least $n$ such
that $a_{2n,
|\sigma|} \in \cnt(\sigma)$ and $a_{2n+1,|\sigma|} \notin \cnt(\sigma)$.  
If $a_0 \notin \cnt(\sigma)$ and $\cnt(\sigma) \neq \emptyset$, then
$M$ conjectures 
$[\min(\cnt(\sigma))]$.  If $\cnt(\sigma) = \emptyset$, then $M$
outputs a default 
index, say $A_1$.  An explanatory learner for $\cL_2$ can be
constructed similarly.

\medskip
\noindent
Now assume, by way of contradiction, that $\cL_1 \cup \cL_2$ were vacillatorily 
learnable by some learner $N$.  Note that a slight variant of Proposition 
\ref{prop:lockingsequence} applies to vacillatory learners: if $N$ vacillatorily
learns a language $L$, then there exist a sequence $\sigma \in (L \cup
\{\#\})^*$
and a finite set
$F$ of indices such that $\cnt(\sigma) \subseteq L$ and for all $\tau
\in (L \cup 
\{\#\})^*$, $N(\sigma\circ\tau) \in F$. 
In particular, there exist such a sequence $\sigma_0 \in ([a_0] \cup
\{\#\})^*$ and such a finite set $F_0$ of indices for $N$ on 
$L = [a_0]$.
By the choice of $\sigma_0$ and $F_0$ and the fact that $N$
vacillatorily learns 
all sets of the form $[a_0,a_k]$, where $k > 0$, it follows that the r.e.\ set
$$
\{x: \exists\tau\in[a_0,x]^*\,[N(\sigma_0\circ\tau) \notin F_0]\}
$$ 
is an $\eta$-infinite r.e.\ subset of $\N \sm [a_0]$.  
But this contradicts the assumption that there is no $\eta$-infinite r.e.\ set 
different from $\N$.  Thus neither explanatory nor vacillatory learning is 
closed under union with respect to $\eta$. 

Case 2: There is an $\eta$-infinite r.e.\ set that is different
from $\N$.
Let $B$ be any $\eta$-infinite r.e.\ set that is different from $\N$,
and fix some $k$ such that $a_k \notin B$.
Let $\eta_s$ be the equivalence relation obtained by enumerating 
$\eta$-equivalent pairs of elements of $B$ for $s$ steps and keeping the 
relation closed transitively and reflexively. 
Furthermore, $b_{n,s}$ is the $n$-th element in the order of the enumeration 
satisfying that all elements before it in the enumeration are not $\eta_s$ 
equivalent to $b_{n,s}$. Now set
\begin{equation*}
\begin{split}
\cL_1 &= \{A_{n+k+1}: n \in \N\}, \\
\cL_2 &=  \{\{x: \exists m < n \,\exists s\,[x\, \eta_s\, b_{m,s}]\}:
n \in \N\} \cup \{\N\}.
\end{split}
\end{equation*}  
By Proposition \ref{prop:subascendsuperfamily}, $\cL_1$ is an $\eta$-family.
Furthermore, a proof similar to that of Proposition \ref{prop:ascendsuperfamily}
shows that every uniformly r.e.\ superfamily of 
$\{\{x: \exists m < n \,\exists s\,[x\, \eta_s\, b_{m,s}]\}: n \in
\N\}$ is an $\eta$-family; in 
particular, $\cL_2$ is an $\eta$-family.

\medskip 
\noindent
By Theorem \ref{thm:ascendexnotconf}, $\cL_1$ is explanatorily learnable.
An explanatory learner $M$ for $\cL_2$ works as follows: on input $\sigma$,
if $a_k \in \cnt(\sigma)$, then $M$ outputs a canonical index for $\N$;
if $a_k \notin \cnt(\sigma)$, then $M$ conjectures $\{x: \exists m <
n\, \exists s\,[x\,\eta_s\, b_{m,s}] \}$ for the least $n$ found such that 
$b_{n,|\sigma|} \notin \cnt(\sigma)$, where $b_{n,|\sigma|}$ is
defined as earlier.
If $M$ is presented with a text for $\N$, then $M$ will always output 
a canonical index for $\N$ after $a_k$ appears in the text.  If $M$ is 
presented with a text $T$ for some set $\{x: \exists m < n\, \exists
s\,[x\,\eta_s\, b_{m,s}]\}$, 
then there is an $s_0$ large enough so that $b_m \in \cnt(T[s_0])$ for
all $m < n$ 
and $b_{m,s_0} = b_m$ for all $m \leq n$; thus $M(T[s'])$ equals some canonical 
index for $\{x: \exists m < n\, \exists s\,[x\,\eta_s\, b_{m,s}]\}$
whenever $s' \geq s_0$.   

\medskip
\noindent
Since $\cL_1 \cup \cL_2$ contains $\N$ and the infinite ascending
chain $\{A_{k+1} 
\subset A_{k+2} \subset \ldots \subset A_{k+i} \subset \ldots\}$
whose union is $\N$,
$\cL_1 \cup \cL_2$ 
is not behaviourally correctly learnable.  Hence, under the present
case assumption,
explanatory (resp.~vacillatory, behaviourally correct) learning is not closed
under union with respect to $\eta$.  (In particular, this proves the
``only if'' 
direction of statement (b).)~\qed
}
\proofofnonunionex

\medskip
\noindent
The next result establishes the non-union theorem for finite learning of
$\eta$-families, where
$\eta$ is any positive equivalence relation such that at least one
finitely learnable 
$\eta$-family exists. It may be worth noting that, in contrast to explanatory 
learnability, there is a positive equivalence relation $\vartheta$ for which no 
$\vartheta$-family is finitely (or even confidently) learnable, as
will be seen in the 
subsequent section.

\begin{thm}\label{thm:nonunionfinite}
Let $\eta$ be any given positive equivalence relation such that at
least one $\eta$-family
is finitely learnable.  Then there are finitely learnable
$\eta$-families $\cL_1$ and $\cL_2$ 
for which $\cL_1 \cup \cL_2$ is an $\eta$-family that is not finitely learnable.
Furthermore, $\cL_2 - \cL_1$ contains a single language.
\end{thm} 

\def\proofofnonunionfinite{
\ownproof
Let $\{B_0,B_1,B_2,\ldots\}$ be a finitely learnable $\eta$-family, as
witnessed by some learner $M$. By a slight variant of Proposition
\ref{prop:lockingsequence} for finite 
learning, there exists for every $i \in \N$ a finite sequence
$\sigma_i \in (B_i \cup 
\{\#\})^*$ and an index $e$ with $W_e = B_i$ such that $M(\tau) = ~?$
for any proper 
prefix $\tau$ of $\sigma_i$ and for all $\sigma \in (B_i \cup
\{\#\})^*$, $M(\sigma_i 
\circ \sigma) = e$. Set $F_i := \cnt(\sigma_i)$.  The family
$\{F_0,F_1,F_2,\ldots\}$ is
\emph{uniformly recursively generable} \cite{LZ93b} in that there is a
total effective 
procedure that, on every input $i$, generates all elements of $F_i$ and stops.  
Furthermore, if there were some $j \neq i$ such that $F_i \subseteq
B_j$, then one could 
extend $\sigma_i$ to a text $T$ for $B_j$ and so $M$ would not
finitely learn $B_j$ on $T$ 
because it outputs an index for $B_i$ on $\sigma_i$.  Thus $F_i$ is a
\emph{characteristic 
sample} \cite{lz-colt-92,Muk:c:92:fin} for $B_i$ in the sense that
$F_i \subseteq B_i$ 
and for all $j \neq i$, $F_i \not\subseteq B_j$.  The family
$\cL_1:=\{[F_0],[F_1],[F_2],
\ldots\}$ is therefore an $\eta$-family.  Note that $\cL_1$ is finitely
learnable: on any given text, a finite learner outputs $?$ until it
identifies the least 
$i$ (if there is any such $i$) such that --- (1) $F_i$ is contained in the
range of the current input and
(2) one has seen at least $i$ data items, including pause symbols, in the text
---
then the learner conjectures $[F_i]$.
Condition (2) is only
delaying the learning and is needed to make the learner recursive.
Furthermore, it is possible to check for any $i$ whether 
or not $F_i$ belongs to the range of the current input because
$\{F_0,F_1,F_2,\ldots\}$ 
is uniformly recursively generable).

Let $\cL_2$ be the family $\{[F_0 \cup F_1]\} \cup (\cL_1 \sm
\{[F_j]: [F_j] \subset [F_0 \cup F_1]\}$.  Since $[F_0 \cup F_1]$ is
$\eta$-finite, there are only finitely 
many indices $j$ such that $[F_j] \subset [F_0 \cup F_1]$ and
therefore $\cL_2$ is a 
finite variant of $\cL_1$.  It follows that $\cL_2$ is also an $\eta$-family. 
Note that $\cL_2$ is finitely learnable: on input $\sigma$, a finite
learner outputs
$?$ until it identifies the least $i$ (if there is any such $i$) such that 
$F_i$ is contained in the range of the current input and either $i \in
\{0,1\}$ or $[F_i] 
\not\subseteq [F_0 \cup F_1]$.  If $i \in \{0,1\}$, then the learner
conjectures $[F_0 \cup 
F_1]$, and if $i > 1$, then it conjectures $[F_i]$.  Suppose the
learner is presented with a 
text for $[F_0 \cup F_1]$.  Then the least $j$ such that $F_j$ is
contained in the range of the 
text and either $j \in \{0,1\}$ or $F_j \not\subseteq [F_0 \cup F_1]$
must be either $0$ or $1$; 
thus the learner will settle on the conjecture $[F_0 \cup F_1]$.
If the learner is presented with a text for some $[F_j]$ with $[F_j]
\not\subseteq [F_0 \cup 
F_1]$, then, since $j \notin \{0,1\}$ and $j$ is the least index $j'$
such that $F_{j'}$ is contained in the range of the text and $[F_{j'}]
\not\subseteq [F_0 \cup F_1]$,
the learner will settle on the conjecture $[F_j]$.

Now consider $\cL_1 \cup \cL_2$; this is an $\eta$-family because it
is the union of
the $\eta$-family $\cL_1$ and $\{[F_0 \cup F_1]\}$.  Since any finite
subset of $[F_0]$ 
is contained in $[F_0 \cup F_1]$, $[F_0]$ does not have a finite
characteristic sample 
-- which, as was pointed out earlier, must exist for every language in
a finitely 
learnable family. Thus $\cL_1 \cup \cL_2$ is not finitely learnable.~\qed
}
\proofofnonunionfinite

\medskip
\noindent
Note that confidently learnable classes (where learners are possibly 
non-partial-recursive) are closed under union 
\cite[Exercise 4.6.2C]{OSW86} 
and that similarly adding single languages does not destroy confident
learnability. The remainder of this section considers a less restrictive
notion of closure under union for learning criteria; this is motivated
by the fact that the union of two $\eta$-families is not necessarily
an $\eta$-family: Letting $\eta$ to be equality-relation and
$A$ be a nonrecursive r.e.\ set,
the $\eta$-families $\{2x: x \notin A\} \cup \{2x,2x+1: x \in A\}$
and $\{2x+1: x \notin A\} \cup \{2x,2x+1: x \in A\}$ satisfy the property
that their union is not an $\eta$-family.

\begin{defn}
(1) Given uniformly
r.e.\ families (with corresponding one-one indexing)
$\cL := \{L_0,\linebreak[3] L_1,\ldots\}$ and $\cH := \{H_0,H_1,\ldots\}$,
say that $\cL$ is \emph{one-one reducible to $\cH$} iff there is a one-one
recursive function $f$ such
that for all $e \in \N$, $L_e = H_{f(e)}$.

(2) An $\eta$-family ${\cal K}$ is the strong union of $\cL$ and $\cH$
iff ${\cal K} = \cL \cup cH$ and there are one-one reductions from
$\cL$ to ${\cal K}$ and from $\cH$ to ${\cal K}$.

(3) For any learning criterion $I$ and any positive 
equivalence relation $\eta$, say that $I$ is \emph{strongly closed
under union with respect to $\eta$} iff for any $I$-learnable 
$\eta$-families $\cL$ and
$\cH$, also every strong union of $\cL$ and $\cH$ is exactly $I$-learnble.
\end{defn}

\begin{thm}\label{thm:strongclosurefinconfexbc}
For any positive equivalence relation $\eta$, confident learning is
strongly closed under union with respect to $\eta$ but neither finite
nor explanatory
learning is strongly closed under union with respect to $\eta$. 
\end{thm}

\def\proofofstrongclosurefinconfexbc{
\ownproof
Suppose $\cL := \{L_0,L_1,L_2,\ldots\}$ and $\cH :=
\{H_0,H_1,H_2,\ldots\}$ are exactly 
confidently learnable $\eta$-families such that $\cL \cup \cH =
\{B_0,B_1,B_2,\ldots\}$
is an $\eta$-family, $\cL$ is one-one reducible to $\cL \cup \cH$ and
$\cH$ is one-one 
reducible to $\cL \cup \cH$.  Let $f$ and $g$ be recursive functions
such that for all 
$n \in \N$, $L_n = B_{f(n)}$ and $H_n = B_{g(n)}$, and let $M_1$ and
$M_2$ be exact 
confident learners for $\cL$ and $\cH$ respectively.
An exact confident learner $N$ for $\cL \cup \cH$ works as follows: on
input $\sigma$,
simulate $M_1$ and $M_2$ and let $d$ and $e$ be their respective
outputs on $\sigma$. 
If $f(d) = g(e)$, then $N$ outputs $f(d)$.  If $f(d) \neq g(e)$, then
$N$ searches for
an $s > |\sigma|$ such that there is a least $x$ with $x \in
B_{f(d),s} \triangle 
B_{g(e),s}$ (such $s$ and $x$ must exist because $B_0,B_1,B_2,\ldots$
is a one-one 
numbering and $f(d) \neq g(e)$).
Set 
$$
N(\sigma) = \left\{\begin{array}{ll}
f(d) & \mbox{if $x \in \cnt(\sigma) \cap B_{f(d),s}$ or $x \notin
\cnt(\sigma)\cup B_{f(d),s}$;} \\
g(e) & \mbox{if $x \in \cnt(\sigma) \cap B_{g(e),s}$ or $x \notin
\cnt(\sigma)\cup B_{g(e),s}$.}
\end{array}\right.
$$
On any text $T$, both $M_1$ and $M_2$ will converge; suppose $M_1$ and
$M_2$ converge 
to $d_0$ and $e_0$ respectively.
If $f(d_0) = g(e_0)$, then $N$ will converge to $f(d_0)$; if $f(d_0)
\neq g(e_0)$, 
then $N$ will converge to $f(d_0)$ (resp.~$g(e_0)$) if the least $x$ such that
$x \in B_{f(d_0)} \triangle B_{g(e_0)}$ satisfies $x \in \cnt(T) \cap
B_{f(d_0)}$ or
$x \notin \cnt(T) \cup B_{f(d_0)}$ (resp.~$x \in \cnt(T) \cap B_{g(e_0)}$ or
$x \notin \cnt(T) \cup B_{g(e_0)}$).  Furthermore, if $T$ is a text for any
$B_{\ell} \in \cL \cup \cH$, then at least one of $M_1$ and $M_2$
converges on $T$ to
a correct index for $B_{\ell}$, so that $\ell \in \{f(d_0),g(e_0)\}$. 
If $f(d_0) = 
g(e_0)$, then $\ell = f(d_0) = g(e_0)$ and so $N$ will output $\ell$
in the limit; if 
$\ell = f(d_0) \neq g(e_0)$, then there is a least $x$ in $B_{f(d_0)} \triangle 
B_{g(e_0)}$ such that either $x \in \cnt(T) \cap B_{f(d_0)}$ or $x
\notin \cnt(T) 
\cup B_{f(d_0)}$, and so $N$ will again output $\ell$ in the limit; a
similar argument 
applies if $\ell = g(e_0) \neq f(d_0)$.  It follows that $N$ exactly
confidently learns 
$\cL \cup \cH$.

For explanatory learning, one notes that the classes $\cL_1$ and $\cL_2$ in both
Cases 1 and 2 of the proof of Theorem \ref{thm:nonunionex} are
$\eta$-families with only finitely much intersection, 
and so if $\cL_1 = \{C_0,C_1,C_2,\ldots\}$ and $\cL_2 =
\{E_0,E_1,E_2,\ldots\}$, then, 
setting $F_{2i} = C_i$ and $F_{2i+1} = E_i$ for all $i \in \N$,
$\{F_0,F_1,F_2,\ldots\}$ 
is a almost a one-one uniformly r.e.\ numbering of $\cL_1 \cup \cL_2$
(finitely many languages may have two indices).
Then $i \mapsto 2i$ and $i \mapsto 2i+1$ are recursive functions witnessing
the one-one reductions of $\cL_1$ to $\cL_1 \cup \cL_2$ and of $\cL_2$ to
$\cL_1 \cup \cL_2$ respectively. 
This can be easily modified to have a 1--1 indexing of $\cL_1 \cup \cL_2$.
Furthermore, 
the classes $\cL_1$ and $\cL_2$ in both Cases 1 and 2 of the proof of Theorem 
\ref{thm:nonunionex} are explanatorily learnable (even exactly).  
Thus explanatory learning is not strongly closed under union.  A
similar argument 
based on the proof of Theorem \ref{thm:nonunionfinite} shows that
finite learning is 
also not strongly closed under union.~\qed
}
\proofofstrongclosurefinconfexbc

\section{Learnability of Families Closed Under Special Positive
Equivalence Relations}

\noindent
So the general results were that for every positive equivalence
relation $\eta$, for each of the following conditions, 
there are $\eta$-families which satisfy it: (a) the family is explanatory
learnable but not confidently learnable; (b) the family is
behaviourally correctly learnable but not vacillatorily learnable;
(c) the family is not behaviourally correctly learnable. The picture
does not provide $\eta$-families which are confidently learnable
and also not separate out the notion of vacillatory learning
from explanatory learning. The first main result of this section is to
construct a positive equivalence relation $\vartheta$ such that
there is no confidently learnable $\vartheta$-family and furthermore
all vacillatorily learnable $\vartheta$-families are explanatory learnable.
\ifx\versy\fully
Thus one cannot separate for all $\eta$ the notions of vacillatory
and explanatory learning and one also cannot show that every
$\eta$ has a confidently learnable $\eta$-family.
\fi
The second main result shows that there is a positive equivalence
relation $\zeta$ for which there are confidently learnable 
$\zeta$-families but no finitely learnable $\zeta$-families.

\begin{thm}\label{thm:varthetaproperties}
 There is a positive equivalence relation $\vartheta$ satisfying:
 \begin{enumerate}[(1)]
  \item There is only one 
$\vartheta$-infinite r.e.\ set, namely $\mathbb{N}$.
  \item Every $\vartheta$-family contains an infinite ascending chain
$B_0 \subset B_1 \subset \cdots$ of $\vartheta$-finite sets whose
union is $\mathbb{N}$.  In particular, no $\vartheta$-family is
confidently learnable and every behaviorally correctly learnable
$\vartheta$-family consists only of $\vartheta$-finite languages.
  \item Every vacillatory learnable $\vartheta$-family is
explanatorily learnable.
 \end{enumerate}
\end{thm}

\def\proofofvarthetaproperties{
\ownproof
Recall that $W_n$ denotes the $n$-th recursively enumerable set, and
$W_{n,s}$ is the finite part of $W_n$ that is enumerated before the
given stage $s \in \mathbb{N}$. Recall also the definitions of $a_m
\in \mathbb{N}$ and $A_n = [a_0,a_1,\ldots,a_{n-1}]$.

\medskip
\noindent
{\bf Construction} of $\vartheta$:
Initially, let $a_{m,0} = m$ for all $m$ --- that is, to start, 
$\vartheta$ consists wholly of singleton equivalence classes.
Then, for each $s=0,1,\ldots$ in turn, search for a triple $n,k,\ell
< s$ satisfying:
\begin{itemize}
    \item $n < \ell$ and $k < \ell$;
    \item $[a_{k,s}]_s \cap W_{n,s} = \emptyset$; and
    \item $[a_{\ell,s}]_s \cap W_{n,s} \neq \emptyset$.
\end{itemize}
If such a triple $n,k,\ell$ exists, then select one with $\ell$ as small 
as possible, merge the classes $[a_k]$ and $[a_{\ell}]$ together, 
that is, let $a_{k,s} \vartheta a_{\ell,s}$ (along with any of
its implications based on $\vartheta$ being equivalence relation), and then
set  $a_{\ell',s+1}=a_{\ell',s}$, for $\ell' < \ell$ and
$a_{\ell',s+1}=a_{\ell'+1,s+1}$, for $\ell' \geq \ell$,
so that each $a_{m,s+1}$ is again the least element of the $m$-th class. 
If no such triple exists, then $a_{\ell',s+1}=a_{\ell',s}$ for all $\ell'$.
Then move on to the next value of $s$.
This completes the construction.

\begin{clm}
The equivalence relation $\vartheta$ thus produced has infinitely many
classes (as demanded in the last
paragraph of Section \ref{sec:not}).
\end{clm}

\noindent
{\it Proof:} 
It is enough to verify that each $m 
\in \mathbb{N}$ serves the role of $\ell$ only finitely many times, and hence 
that each $a_m$ reaches some limiting value. This is easily seen by induction 
on $m$: after each $a_k$, $k<m$ has reached its limiting value, $a_{m,\cdot}$,
changes its value for each pair $n,k<m$ at most once. Thus, each $a_m=\lim_{s 
\to \infty} a_{m,s}$, reaches its limiting value.

\begin{clm}\label{clm2}
For each $W_n$ which is $\vartheta$-closed, either: $W_n \subseteq
A_n$, or $W_n = A_m$ for some $m \geq n$, or $W_n = \mathbb{N}$. In
particular, $\vartheta$ satisfies part (1) of the theorem.
\end{clm}

\noindent
{\it Proof:} If a $\vartheta$-closed $W_n$ contains (the limiting
value of) a given $a_\ell$ with $\ell \geq n$, then by construction
$W_n$ also contains $\{a_0,\ldots,a_{\ell-1}\}$. Hence if $W_n$ is
$\vartheta$-closed and not contained in $A_n$, we have either $W_n =
A_m$ for some $m$, or $W_n = \mathbb{N}$.

\begin{clm}\label{clm3} If $\mathcal{L}$ is a $\vartheta$-family then
for every $n$ there are infinitely many $L \in \mathcal{L}$ such that
$A_n \subseteq L$.
\end{clm}

\noindent
{\it Proof:} By induction on $n$. The base case, where $A_0 =
\emptyset$, is trivial. For $n \geq 1$, let $U = \bigcup \{L \in
\mathcal{L} : A_{n-1} \subseteq L\}$. Clearly $U$ is
$\vartheta$-closed and r.e.; and by the induction hypothesis $U$ has
infinitely many different summands $L$, so $U$ is $\vartheta$-infinite
by a pigeonhole argument. It follows by Claim~\ref{clm2}
that $U = \mathbb{N}$, and in particular $a_{n-1} \in U$. Thus
$a_{n-1}$ is an element of one of $U$'s summands $L_0$, so that $A_n
\subseteq L_0$. Repeat the construction with $L_0$ omitted from
$\mathcal{L}$ to get a second such $L_1$, then $L_2$, and so on.

\medskip
\noindent
Note that by Proposition \ref{prop:natetainf} and part (1) of the
theorem, every behaviourally correctly
learnable $\vartheta$-family contains only $\vartheta$-finite sets. 
Thus, in light of Claims~\ref{clm2} and \ref{clm3},
part (2) of the theorem is true.    

\begin{clm}Every vacillatory learnable $\vartheta$-family is
explanatory learnable.
\end{clm}

\noindent
{\it Proof:} Suppose $\mathcal{L}$ is a $\vartheta$-family with vacillatory 
learner $M$; we know from part (2) of the Theorem (that is, from the previous 
claim) that $\mathcal{L}$ consists only of $\vartheta$-finite sets. Given an
enumeration of a language $L \in \mathcal{L}$, the learner $M$ outputs at
most finitely many indices, say with maximum $n^*$; and given this $n^*$, we 
know from an earlier claim that either $L \subseteq A_{n^*}$, or 
$L = A_m$ for some $m \geq n^*$. So the new explanatory learner $N$ works by 
watching for the largest $n$ output by $M$; then $N$ watches to see if $a_{n}$ 
is a member of $L$ and if so, $N$ conjectures that $L = A_m$ for the least
$m$ such that $a_{m}$ does not belong to the input text;
otherwise $N$ conjectures that $L$ is an appropriate subset of 
$A_{n}$ (which consists of the elements 
$a_i$, $i<n$, which have appeared in the input text).
It is easy to verify that $N$ explanatorily learns $\mathcal{L}$.
\hspace*{\fill}\qed
}
\proofofvarthetaproperties

\medskip
\noindent
According to Theorem \ref{thm:finconfsepallper}, for every positive
equivalence relation $\eta$ such that there is at least one finitely
learnable $\eta$-family, there is also an $\eta$-family that is 
confidently but not finitely learnable.  The next main result 
complements this theorem by showing that there is a positive 
equivalence relation $\zeta$ for which no finitely learnable 
$\zeta$-family exists even though there are confidently learnable
$\zeta$-families. 

\begin{desc} \label{de:zeta} \rm
One defines a positive equivalence relation $\zeta$ using a
dense simple set $Z$ with $0 \notin Z$ as below; recall for this
that a set is dense simple iff it is recursively enumerable, coinfinite
and the sequence $a_0,a_1,\ldots$ of its non-elements in ascending order
grows faster than every recursive function. It is known that such
sets $Z$ exist \cite{Od89}.
\sp
Now one defines that $x \, \zeta \, y$ iff
there is an $n$ with $a_n \leq \min\{x,y\} \leq \max\{x,y\} < a_{n+1}$.
This relation is positive, as $x \, \zeta \, y$ is equivalent to
$$
   \forall z\,[\min\{x,y\} < z \leq \max\{x,y\} \Rightarrow z \in Z]
$$
which is an r.e.\ condition.  Furthermore, in coincidence with the
notation used in this paper, each $a_n$ is the least element of its
equivalence class and the $a_n$ are the ascending limits of the approximations
$a_{n,t}$ which are the non-elements (in ascending order) of the set $Z_t$
of the first $t$ elements enumerated into $Z$, so that $Z_0 = \emptyset$ and
$a_{n,0} = n$.  As $Z$ is coinfinite, there are infinitely many $a_n$'s
and so $\zeta$ induces infinitely many equivalence classes.  
\end{desc}

\begin{thm}\label{thm:confzetanotfinite}
There is a confidently learnable $\zeta$-family but no
finitely learnable $\zeta$-family.
\end{thm}

\def\proofofconfzetanotfinite{
\ownproof
First one defines $B_n = \{x: x \, \zeta \, n \vee x \, \zeta \, a_n \vee
n \leq x \leq a_n\}$. Note that $n \leq a_{n,0} \leq a_n$ and thus
the minimum of $B_n$ is the maximum of all $a_m$ with $a_m \leq n$.
Furthermore, each two different sets $B_n$ and $B_m$ have the different
maxima $a_n$ and $a_m$ (among different $a_i$'s),
respectively, and therefore the so defined
family is an r.e.\ one-one family of $\zeta$-closed sets. So it is
a $\zeta$-family. A confident learner conjectures on an input sequence
$\sigma$, whenever it exists, the least $n$ such that $range(\sigma)
\subseteq B_{n,|\sigma|}$; however, if such a set does not exist
since none of the finitely many sets with $\min(B_{n,|\sigma|}) \leq
\min(range(\sigma))$ contains $range(\sigma)$, then the learner
conjectures $B_0$. Note that all $B_n$ are $\eta$-finite and each number is only
contained in finitely many $B_n$. Thus, whenever there is an $n$ for which
$B_n$ contains all the data in the text then the learner converges
to an index of $B_n$ for the least such $n$ and when no such $n$ exists
then the learner converges to an index of $B_0$. Hence the learner is
confident and learns the family.
\sp
Second let $E_0,E_1,\ldots$ be any $\zeta$-family and assume by way
of contradiction that it is finitely learnable. Then for
each $E_n$ there is a finite subset $C_n \subseteq E_n$
such that the finite learner conjectures some index different from $?$
after having seen the elements $C_n$; the mapping, call it $f$,
from $n$ to an explicit list of the set $C_n$ is recursive and so
is $g$ defined by $g(n) = \max\{\max(C_m): m \leq 2^n\}$.
Also the function $g$ is recursive and by choice of $Z$, there is an
$n$ with $g(n) < a_n$. It follows that the $2^n+1$ sets $E_m$
with the indices satisfying $m \leq 2^n$ are all
learnt based on data which is contained in 
$[a_0$, $a_1$, $\ldots$, $a_{n-1}]$ and therefore there are two distinct numbers
$m,k \leq 2^n$ such that the $\zeta$-closure of $C_m$ and $C_k$ are the
same sets. It follows that both sets $C_m$ and $C_k$
must be subsets of both $E_m$ and $E_k$ and so the finite learner must
err when learning the sets $E_m$ and $E_k$. Hence the given $\zeta$-family
is not finitely learnable.~\qed
}
\proofofconfzetanotfinite

\section{Weakly Confident and Class-Preserving Learning}

\noindent
As Osherson, Stob and Weinstein \cite{OSW86} had already observed, 
confidence is a real restriction on explanatory or behaviourally correct 
learning; in particular, even the relatively simple family of all finite sets 
is not confidently behaviourally correctly learnable.  
In the present section, we examine a variant of confident learning
known as \emph{weak confident learning}, where the learner's 
sequence of hypotheses is required to converge only on each text 
whose range is contained in some language to be learnt.
Such a learning property may be desirable in situations where a 
text for some target language may have missing data due to noise 
\cite{Case00}.  One can see, for example, that since the family of all finite
sets is closed under taking subsets, it is weakly confidently explanatorily 
learnable.

We also study the effect of requiring \emph{class-preservingness} 
of the learner's hypothesis space on the strength of confident as well as 
weakly confident learning. 
Our first observation is that there are positive equivalence relations
$\eta$ for which class-preserving learning is indeed a restriction on
confidence with respect to the class of $\eta$-families.

\begin{thm}\label{thm:existsetanotconfclspresv}
There is a positive equivalence relation $\eta$ for which there is 
a $\eta$-finite family that is confidently learnable but not class-preservingly
confidently learnable. 
\end{thm}

\def\proofofexistsetanotconfclspresv{
\ownproof
Take $\eta$ to be the positive equivalence relation $x\, \eta\, y
\Leftrightarrow x = y$.
Now consider the class $\cC$ comprising all $2$-element sets $\{x,y\}$
with $x,y \in \N$,
$x \neq y$, and all singletons $\{z\}$ with $z \in K'$.

Define a confident learner $M$ as follows.  On input $\sigma$, if
$|\cnt(\sigma)| 
\leq 2$, then $M$ conjectures $\cnt(\sigma)$; if $|\cnt(\sigma)| > 2$, then $M$ 
outputs a default index.  Then $M$ will converge on any text $T$
containing at most
two elements to a canonical index for $\cnt(T)$, and so it explanatorily learns
every language in $\cC$; if, on the other hand, $\cnt(T)$ contains more than
two elements, $M$ will converge to a default index.  Hence $M$
confidently learns
$\cC$.  

Assume that $\cC$ has a class-preserving confident learner $N$.  Then, given 
any $n\in\N$, one can check via oracle $K$ whether or not $n \in K'$
as follows: 
feed $N$ with the text $n \circ n \circ n \circ \ldots$ until $N$ converges to 
some index $e$ (this can be checked using oracle $K$).  Then $x \in K'$
holds iff $W_e = \{n\}$ (the latter condition can also be checked
using oracle $K$), 
and so $K' \leq_T K$, but this is impossible.~\qed
}
\proofofexistsetanotconfclspresv

\medskip
\noindent
The remainder of this section will consider a less stringent variant of 
confidence called \emph{weak confidence}.  It will be shown that for 
every positive equivalence relation $\eta$, there is an $\eta$-family
witnessing the separation between explanatory learning and weak confident
learning.   

\begin{defn}\label{defn:weakconf}
For any given uniformly r.e.\ family $\cC = \{L_0,L_1,\ldots\}$, a
text $T$ is said 
to be \emph{consistent with $\cC$} iff there is an $e$ such that
$\cnt(T) \subseteq 
L_e$.  A learner $M$ is \emph{weakly confident} iff $M$ converges on
every text that
is consistent with the target class $\cC$. 
\end{defn}

\medskip
\noindent
We first observe that for any positive equivalence relation $\eta$, 
weak confidence is indeed a less stringent learning constraint than confidence 
with respect to the class of $\eta$-families.

\begin{thm}\label{thm:sepweakconfconf}
Let $\eta$ be any given positive equivalence relation.  Then $\cA_{\eta}$ 
is weakly confidently learnable but not confidently learnable.
\end{thm} 

\def\proofofsepweakconfconf{
\ownproof
It has been proven in Theorem \ref{thm:ascendexnotconf} that the ascending
family is not confidently learnable.  Furthermore, the explanatory
learner constructed in the proof of Theorem \ref{thm:ascendexnotconf}
is in fact weakly confident.  To see this, note that for any text $T$ such 
that $\cnt(T) \subseteq A_e$ for some $e \in \N$, there is a least $a_n$
such that $a_n \notin \cnt(T)$, and so the explanatory learner constructed 
in the proof of Theorem \ref{thm:ascendexnotconf} will, on $T$, converge to 
an index for $A_n$.~\qed
}
\proofofsepweakconfconf

\medskip
\noindent
The next result separates weak confident learning from its class-preserving
variant. 

\begin{thm}\label{thm:weakconfnotclasspresv}
Let $\eta$ be any given positive equivalence relation.
The class $\cC$ consisting of all members of $\cA_{\eta}$ and all $[a_{n+1}]$ 
with $n \in K'$ is an $\eta$-family that is weakly confidently learnable 
but not class-preservingly weakly confidently learnable. 
\end{thm}

\def\proofofweakconfnotclasspresv{
\ownproof
Note that $\cC$ has a uniformly r.e.\ one-one numbering.
First, a uniformly r.e.\ numbering of $\cC$ may be obtained as follows:
for all $n,s,t \in \N$, set 
\begin{eqnarray*}
L_{\spn{2n,s,t}} & = & A_n, \\
L_{\spn{2n+1,s,t}} & = & \left\{\begin{array}{ll}
[a_{n+1}] & \mbox{if $\forall s' \geq s\,\forall t' \geq
t\,[\varphi^{K_{s'}}_{n,t'}(n)\downarrow]$;} \\
A_{n+2} & \mbox{otherwise.}\end{array}\right. 
\end{eqnarray*}
Thus $\cC$ is a uniformly r.e.\ superfamily of $\cA_{\eta}$,
and so by Proposition \ref{prop:ascendsuperfamily}, $\cC$ is also an 
$\eta$-family.

Let $M$ be a learner that works as follows on input $\sigma$.
If $a_{0} \in \cnt(\sigma)$ and $n$ is the least number for which 
$a_{n,|\sigma|} \notin \cnt(\sigma)$, $M$ outputs a canonical index for $A_n$; 
if $a_{0} \notin \cnt(\sigma)$ and $\cnt(\sigma) \neq \emptyset$, 
then $M$ conjectures $[\min(\cnt(\sigma))]$; if $\cnt(\sigma) = \emptyset$, 
then $M$ just outputs a default index.  Then $M$ will make infinitely many 
mind changes on a given text $T$ only if, for all $n \in \N$, $a_n$ appears 
in $T$; thus, since no language in $\cC$ contains $\{a_n: n \in \N\}$, $M$ 
will converge on every text for any $L \in \cC$.  Furthermore, given any text 
$T$ for $A_n$, where $n \in \N$, there is an $s$ large enough so that 
$a_{n,t}=a_n$, $a_m \in \cnt(T[t])$
and $a_{m,t} = a_m$ whenever $t \geq s$ and $m < n$, and
so $M(T[t])$ is equal to a canonical index for $A_n$ whenever $t \geq s$.
Similarly, given any text $T'$ for $[a_{n+1}]$, where $n \in \N$, $a_0$
will not appear in $T'$ and so $M$ will converge to an index for $[x]$,
where $x$ is the least element of $[a_{n+1}]$ to appear in $T'$.  Thus
$M$ is a weakly confident learner of~$\cC$.

Assume, by way of contradiction, that $\cC$ has a class-preserving weakly
confident learner $N$.  Then, given any $n\in\N$, one can check via oracle $K$
whether or not $n \in K'$ as follows.  Feed $N$ with the text $a_n \circ a_n
\circ a_n\circ\ldots$ until $N$ converges to an index (since $N$ is weakly
confident and $a_n \in A_{n+1} \in \cC$, $N$ must converge on this
text), say $e$.  
Then $n \in K'$ holds iff $a_{n+1} \in W_e \wedge a_0 \notin W_e$;
the latter condition can be checked using oracle $K$.  This implies 
$K' \leq_T K$, a contradiction.~\qed
}
\proofofweakconfnotclasspresv

\medskip
\noindent
The next two results clarify the relationship between explanatory learning
and weakly confident learning of one-one $\eta$-closed families for any given 
positive equivalence relation $\eta$.

\begin{thm}\label{thm:exweakconfa}
Let $\eta$ be any given positive equivalence relation.  If there are at least 
two $\eta$-infinite r.e.\ sets, then there is an $\eta$-family that is 
explanatorily learnable but not weakly confidently learnable.
\end{thm}

\def\proofofexweakconfa{
\ownproof
Let $B$ be any $\eta$-infinite r.e.\ set such that $B \neq \N$.  Fix some
$b \notin B$.  
Let $\eta_s$ be the equivalence relation obtained by enumerating 
$\eta$-equivalent pairs of elements of $B$ for $s$ steps and keeping the 
relation closed transitively and reflexively. 
Furthermore, $b_{n,s}$ is the $n$-th element in the order of the enumeration 
satisfying that all elements before it in the enumeration are not $\eta_s$ 
equivalent to $b_{n,s}$.
Now set
\begin{equation*}
\begin{split}
B_0 &= \N, \\
B_{n+1} &= \{x: \exists m < n \,\exists s\,[x\, \eta_s\, b_{m,s}]\}
\end{split}
\end{equation*}
for all $n \in \N$.
As was shown in the proof of Theorem \ref{thm:nonunionex}, 
$\cC:= \{B_n: n \in \N\}$ is explanatorily learnable.
On the other hand, $\cC$ is not weakly confidently learnable.
For, if $N$ were a weakly confidently learner for $\cC$, then one
could build a text $T'$ that is the limit of sequences $\sigma_0\circ\sigma_1
\circ\ldots\circ\sigma_s$ of strings, where, for all $n$, $\sigma_n$ is the 
first string in $B_{n+1}^*$ found such that $N$ on input $\sigma_0\circ\ldots
\circ\sigma_n$ outputs an r.e.\ index for $B_{n+1}$; note that since $B_{n+i} 
\subset B_{n+i+1}$ for all $i$, such strings $\sigma_0,\sigma_1,\ldots$ must 
exist.  Then $T'$ is a text consistent with $\cC$ (since $\N \in \cC$) on 
which $N$ makes infinitely many mind changes, which contradicts the fact that 
$N$ is weakly confident.~\qed
}
\proofofexweakconfa

\begin{thm}\label{thm:exweakconfb}
Let $\eta$ be any given positive equivalence relation.  If $\N$ is the only
$\eta$-infinite r.e.\ set, then there is an $\eta$-family that is
explanatorily learnable but not weakly confidently learnable.
\end{thm} 

\def\proofofexweakconfb{
\ownproof
Let $\cC$ be the class containing all $\eta$-closed sets $A_{n+1}$
and $[a_1,a_{n+2}]$ for $n \in \N$.  An explanatory learner $M$ may work 
as follows.  On input $\sigma$, $M$ checks whether or not 
$a_0 \in \cnt(\sigma)$.  If $a_0 \in \cnt(\sigma)$, then $M$ 
determines the least $n$ such that $a_{n+1,|\sigma|} \notin \cnt(\sigma)$
and outputs a canonical index for $A_{n+1}$.
If $a_0 \notin \cnt(\sigma)$, then $M$ determines the least $n$ such that 
$a_{n+2,|\sigma|} \in \cnt(\sigma)$ (if such an $n$ exists) and outputs a 
canonical index for $[a_1,a_{n+2}]$; if no such $n$ exists, then $M$ just 
outputs any default index.

Now assume for the sake of a contradiction that $\cC$ has a weakly confident
learner $N$.  Since $[a_1] \subseteq [a_1,a_{n+2}] \in \cC$ for all $n \in \N$,
$N$ must (by the property of weak confidence) converge on any text for
$[a_1]$.  
In particular, there exists a stabilising sequence $\sigma$ for
$[a_1]$, that is, 
a sequence $\sigma \in [a_1]^*$ such that for all $\tau \in [a_1]^*$,
$N(\sigma) 
= N(\sigma\tau)$ (see Section \ref{sec:learn}).  Set 
$A := \{x: \exists\tau\in([a_1,x]^*\,[N(\sigma) \neq N(\sigma\tau)]\}$.  
Since $N$ explanatorily learns $[a_1,a_{n+2}]$ for all $n \in \N$, $A$ must 
contain $a_{n+2}$ for all but at most one $n \in \N$.  On the other
hand, for all $x \in [a_1]$, 
$x \notin A$ by the choice of $\sigma$.  Thus, taking $B = \bigcup_{x
\in A}[x]$, 
$B$ is an $\eta$-infinite r.e.\ set that is not equal to $\N$, contrary to the 
hypothesis of the theorem.~\qed
}
\proofofexweakconfb

\section{Monotonic, Conservative and Class-Preserving Learning}

\noindent
In this section, we turn out attention to \emph{monotonic} and
\emph{conservative} learning of $\eta$-families for any given positive 
equivalence relation $\eta$.
Monotonicity and conservativeness are fairly natural learning constraints
that have been well-studied in the context of uniformly recursive families 
\cite{LZ93b} as well as more general classes of r.e.\ sets \cite{KS16}.
We first revisit the ascending family, which was briefly studied in Section 
\ref{sec:allposeq}, and prove that this class is in fact conservatively
explanatorily learnable for every positive equivalence relation.

\begin{thm}\label{thm:ascendconservex}
For every positive equivalence relation $\eta$ such that $[x]$ is infinite
for each $x$, $\cA_{\eta}$ is conservatively explanatorily learnable.
\end{thm}

\def\proofofascendconservex{
\ownproof
It has already been shown in Theorem \ref{thm:ascendexnotconf} that
this family is explanatorily learnable.
One may modify the explanatory learner constructed in the proof 
of this theorem to obtain a conservative
explanatory learner $N$ as follows: If the original learner's
conjecture at step $t$ is 
$A_{n_t}$ and there is a least $s \geq t$ such that for some $m \leq n_t$, 
$a_{m,s} \neq a_{m,s+1}$, then $N$ only enumerates elements
of $A_{n_t}$ until step $s-1$; in particular, $N$ will not enumerate 
any $a_{i,t}$ with $i \geq n_t$ into its conjecture.  
Thus if there is a first subsequent step 
$t' > t$ such that $a_{m,t'} \neq a_{m,t'-1}$ for some $m \leq n_t$, 
$N$ can check in finitely many steps whether or not the range of the 
current input contains some element not belonging to its prior conjecture; 
if so, then it updates its conjecture to $A_{n_{t'}}$; otherwise, it 
repeats its prior conjecture.  If there is a first subsequent step $t' > t$ 
at which the text enumerates $a_{n_t,t}$ but $a_{n_t,t}\, \eta\, a_{m,t}$ for 
some $m < n_t$, then by construction $N$ would not have enumerated 
$a_{n_t,t}$ in its prior conjecture and thus it is conservative in this case 
as well.~\qed
}
\proofofascendconservex

\medskip
\noindent
The next result shows that for any positive equivalence relation $\eta$,
there is an $\eta$-subfamily of the ascending family for $\eta$ that
has no class-preserving conservatively behaviourally correct learner.

\begin{thm}\label{thm:ascendnotclsprevconsvbc}
For every positive equivalence relation $\eta$, there is an $\eta$-family 
contained in $\cA_{\eta}$ that is neither class-preservingly conservatively
behaviourally correctly learnable nor class-preservingly strongly monotonically
behaviourally correctly learnable.
\end{thm}

\def\proofofascendnotclsprevconsvbc{
\ownproof
Let $\cL$ contain all $A_{2n}$ with $n \in \N$ and all $A_{2n+1}$
where $W_n$ is finite. 
This class is r.e., as one can choose $L_{\langle n,0\rangle} = A_{2n}$ and
$L_{\langle n,m+1\rangle} = A_{2n+1}$ 
in the case that $|W_n| \leq m$ and $A_{2n+2}$ in the case that $|W_n| > m$. 
This enumeration is clearly r.e.\ and one can conclude by Proposition 
\ref{prop:subascendsuperfamily} that there is 
an $\eta$-family for this class. Now assume that there is a
class-preserving conservatively 
behaviourally correct learner for the family. If this learner does not
output on any sequence 
from $A_{2n+1}^*$ a hypothesis containing $a_{2n}$ (which can be
computed using $K$) 
and which is in $A_{2n+1}\setminus A_{2n}$, then the learner does not
learn $A_{2n+1}$ 
and thus $W_n$ is infinite. If this learner outputs on some sequence
from $A_{2n+1}^*$ a 
hypothesis containing $a_{2n}$ then one tests with $K$ whether the hypothesis 
also contains $a_{2n+1}$. If so, again $W_n$ is infinite, as due to
class-preservingness of 
the learner, the only possible hypotheses issued are $A_{2n+2}$ or a
superset of this set 
and then the learner cannot learn $A_{2n+1}$ by conservativeness. If
not then the only 
hypothesis possible is $A_{2n+1}$ by class-preservingness of the
learner and therefore 
$A_{2n+1}$ is in the class and $W_n$ is finite. Thus the existence of
a class-preserving 
conservatively behaviourally correct learner for $\cL$ allows to
decide with the halting problem 
whether a given set $W_n$ is finite or infinite and that is
impossible; thus such a learner 
cannot exist.  Finally, we note that this proof by contradiction also
works if one assumes 
that the learner is class-preservingly strongly monotonically
behaviourally correct.~\qed
}
\proofofascendnotclsprevconsvbc

\begin{thm}\label{thm:ascendnotclspresvconsvex}
For every positive equivalence relation $\eta$, there is an $\eta$-subfamily of 
$\cA_{\eta}$ that is conservatively explanatorily learnable but
not class-preservingly conservatively explanatorily learnable.
\end{thm}

\ownproof
Let $f$ be a limit recursive function approximable from below that dominates 
all recursive functions (for an example of such a function $f$, see the proof 
of \cite[Lemma 7.3(a)]{RC94}); without loss of generality, assume that
$f$ is strictly monotonically increasing.
Define, for all $n \in \N$, 
$$
B_n := \{x: \exists m\leq f(n)\,[x\, \eta\, a_m]\}.
$$
By Theorem \ref{thm:ascendexnotconf}, $\cC = \{B_n: n \in \N\}$ is
conservatively
explanatorily learnable.  Assume that $M$ were a class-preserving conservative
explanatory learner for $\cC$.  Let $g$ be a recursive function such
that for all 
$n\in\N$, $g(n)$ is one more than the maximum of the range of the
first $\sigma \in
\N^*$ found 
for which $M$ makes $n$ mind changes on $\sigma$ and when $M(\tau_1)
\neq M(\tau_2)$ 
and $M(\tau_2) \neq M(\tau_3)$ for prefixes $\tau_1,\tau_2,\tau_3$ of $\sigma$
with $\tau_1 \subseteq \tau_2 \subseteq \tau_3$,
then $\cnt(\tau_1) \subseteq W_{M(\tau_3)}$.
Since $f$ dominates all recursive 
functions, there is some $n$ such that $g(3n+3) \leq f(n)$.
Note that all hypotheses that $M$ issues on prefixes of $\sigma$ are
based on subsets of $B_n$.  There are more than $n$ different hypotheses output
by $M$ on prefixes of $\sigma$ and each must be equal to some $B_m$ with
$m < n$, which is impossible.~\qed

\medskip
\noindent
The next theorem considers the question: given any positive equivalence relation
$\eta$, can one always find an $\eta$-family witnessing the separation
of monotonic behaviourally correct and explanatory learning?  It turns out that
this is indeed the case iff there are at least two distinct
$\eta$-infinite r.e.\ sets.  

\begin{thm}\label{thm:clspresvmonexnotsmon}
Let $\eta$ be any given positive equivalence relation. The following
conditions are equivalent.
\begin{enumerate}[(1)]
\item There is an $\eta$-family that is explanatorily learnable but not
monotonically behaviourally correctly learnable.
\item There is an $\eta$-family that is class-preservingly weakly monotonically 
explanatorily learnable but not monotonically behaviourally correctly learnable.
\item There is an $\eta$-infinite r.e.\ set that is different from $\N$.
\end{enumerate}  
\end{thm}

\def\proofofclspresvmonexnotsmon{
\ownproof
It is immediate that $(2) \Rightarrow (1)$.  It will be shown that (1)
$\Rightarrow$ 
(3) $\Rightarrow$ (2). 

(1) $\Rightarrow$ (3). Suppose $\N$ is the only $\eta$-infinite r.e.\ set.
Consider any $\eta$-family $\cC$ that is explanatorily learnable, 
as witnessed by some learner $P$.
First, Proposition \ref{prop:natetainf} implies that for all
$L \in \cC$, $L$
is the union of finitely many equivalence classes of $\eta$.
Let $Q$ be a learner such that on input $\sigma$, 
$Q$ conjectures $\bigcup_{x \in \cnt(\sigma)} [x]$.  
Note that $Q$ is strongly monotonic by construction.
Moreover, given any $L \in \cC$, $Q$ on any text for $L$ will 
conjecture $L$ in the limit because $L$ is composed of 
only finitely many equivalence classes.

(3) $\Rightarrow$ (2). Suppose there is an $\eta$-infinite r.e.\ set $B_0$ 
such that for some $k \in \N$, $a_k \notin B_0$.  Let $\cC$ be the
$\eta$-family 
comprising $B_0$ and $B_{n+1} := A_{n+k+1} = \bigcup_{m \leq n+k}
[a_m]$ for all $n \in 
\N$.

The following learner $M$ explanatorily learns $\cC$ and is both weakly
monotonic and class-preserving.  On input $\sigma$:

Case 1: $a_k \notin \cnt(\sigma)$.  $M$ conjectures $B_0$.

Case 2: $a_k \in \cnt(\sigma)$.

Case 2.1: There is an $n \in \N$ such that $a_{n+k+1,|\sigma|} 
\notin \cnt(\sigma)$ and $a_{m,|\sigma|}$ $\in \cnt(\sigma)$ whenever
$m \leq n+k$.  $M$ outputs an index $e$ such that
$$
W_e = \left\{\begin{array}{ll}
\bigcup_{m \leq n+k} [a_{m,|\sigma|}] & \mbox{if }\{a_m: m \leq k\} 
\ \subseteq  \\
& \{a_{m,|\sigma|}:m \leq n+k\}; \\
B_1 & \mbox{otherwise.}\end{array}\right.
$$ 
Case 2.2: Neither Case 1 nor Case 2.1.  $M$ conjectures $B_1$.

\medskip
\noindent
Note that $M$ always outputs an index for $B_0$ or for a set 
$\bigcup_{m \leq n}[a_m]$ with $n \geq k$, and hence it is 
class-preserving. 
By Case 1 and the fact that $a_k \notin B_0$, $M$ on any text for $B_0$ 
will always conjecture $B_0$.  In particular, $M$ on any text for $B_0$
strongly monotonically explanatorily learns $B_0$.  If $M$ is presented 
with a text $T$ for some $B_{n+1}$, then there is an $s$ large enough so 
that $\{a_m: m \leq n+k\} \subseteq \cnt(T[s])$ and for all $t \geq s$
and $m \leq n+k+1$, $a_{m,t} = a_m$.  By Case 2, for all $t \geq s$, 
$M$ on $T[t]$ will conjecture $B_{n+1}$.  Hence $M$ is a 
class-preserving explanatory learner of $\cC$.  To see that $M$ is also
weakly monotonic on every text for any set $B_{n+1}$, let $\tau_1$ and 
$\tau_2$ be any two sequences in $(B_{n+1} \cup \{\#\})^*$ such that
$\tau_2$ extends $\tau_1$, and consider the following case distinction.

Case i: $a_k \notin \cnt(\tau_2)$.  Then $M$ conjectures $B_0$ on 
both $\tau_1$ and $\tau_2$.

Case ii: $a_k \notin \cnt(\tau_1)$ and $a_k \in \cnt(\tau_2)$.
Since $M$ on $\tau_1$ will conjecture $B_0$ but $a_k \notin B_0$,
the weak monotonicity condition is satisfied vacuously. 

Case iii: $a_k \in \cnt(\tau_1)$.
Suppose $M$ on $\tau_1$ conjectures $B_1$.  As $a_k \in \cnt(\tau_2)$, 
$M$ on $\tau_2$ must output a set $B_i$ with $i \geq 1$, and therefore
$W_{M(\tau_1)} \subseteq W_{M(\tau_2)}$.
Suppose $M$ on $\tau_1$ conjectures some set $B_{n+1}$ with $n \geq 1$.
Then there is some $n' \geq n$ and subset $\{i_0,\ldots,i_{n+k}\}$ of 
$\{0,\ldots,n'+k\}$ with $i_0 < \ldots < i_{n+k}$ such that
\begin{enumerate}[(1)]
\item For all $j \in \{0,\ldots,n+k\}$, $a_{i_j,|\tau_1|} = a_j$;
\item $\{a_{m,|\tau_1|}: m \leq n'+k\} \subseteq \cnt(\tau_1)$.
\end{enumerate}
Note that for all $s,t \in \N$ with $s \leq t$,
$\{a_m: m \in \N\} \subseteq \{a_{m,t}: m \in \N\} \subseteq \{a_{m,s}: 
m \in \N\}$.  Thus, there is some $\ell \geq n+k$ such that
$a_{\ell,|\tau_2|} = a_{n+k} = a_{i_{n+k},|\tau_1|}$, from which it follows
that
\begin{equation*}
\begin{split}
\{a_m: m \leq n+k\} & \subseteq \{a_{m,|\tau_2|}: m \leq \ell \} \\
~ & \subseteq \{a_{m,|\tau_1|}: m \leq i_{n+k}\} \\
~ & \subseteq \cnt(\tau_1) \\
~ & \subseteq \cnt(\tau_2).
\end{split}
\end{equation*} 
Consequently, $M$ on $\tau_2$ will conjecture a set containing
$\bigcup_{m \leq n+k}[a_m]$,
and so $W_{M(\tau_1)} \subseteq W_{M(\tau_2)}$.    

\medskip
\noindent
Now consider any behaviourally correct learner $N$ of $\cC$.  It will 
be shown that $N$ cannot be monotonic.  Since $N$ behaviourally correctly 
learns $B_0$, there is some $\sigma \in B_0^*$ such that $W_{N(\sigma)} 
= B_0$.  Pick some $n \in \N$ such that $a_{n+k+1} \in B_0$ and 
$\cnt(\sigma) \subseteq B_{n+1}$;
such an $n$ exists because $\sigma$ is finite and $B_0$ is an
$\eta$-infinite r.e.\ set.  
As $N$ also behaviourally correctly 
learns $B_{n+1}$, there is some $\tau \in B_{n+1}^*$ such that 
$W_{N(\sigma\circ\tau)} = B_{n+1}$.  But $a_{n+k+1} \in (B_0 \cap B_{n+2}) 
\sm (B_{n+1} \cap B_{n+2}) = (W_{N(\sigma)} \cap B_{n+2}) \sm (W_{N(\sigma
\circ\tau)} \cap B_{n+2})$ implies that $W_{N(\sigma)} \cap B_{n+2} 
\not\subseteq W_{N(\sigma\circ\tau)} \cap B_{n+2}$, and since 
$\cnt(\sigma\tau) \subseteq B_{n+2} \in \cC$, it may be concluded 
that $N$ does not learn $B_{n+2}$ monotonically --  a contradiction.~\qed
}
\proofofclspresvmonexnotsmon

\section{Conclusion}

\noindent
The present work studied how the relations between the most basic
inference criteria for learning from text are impacted when the only
classes to be considered for learning are uniformly r.e.\ one-one
families of sets which are closed under a given positive equivalence
relation $\eta$. One considers the chain of implications
finitely learnable $\Rightarrow$
confidently learnable $\Rightarrow$ explanatorily learnable $\Rightarrow$
vacillatorily learnable $\Rightarrow$ behaviourally correctly learnable
which is immediate from the definitions. When choosing $\eta$ as the
explicitly constructed $\vartheta$ from 
Theorem~\ref{thm:varthetaproperties},
the implication from explanatorily learnable to vacillatorily learnable
becomes an equivalence and the criterion of confidently learnable becomes
void, that is, no $\vartheta$-family satisfies it.
For the positive equivalence relation $\zeta$ from
Description~\ref{de:zeta}, there is a $\zeta$-family
which is confidently learnable, but none which is finitely learnable.
Furthermore, in the case that a finitely learnable $\eta$-family exists
for some $\eta$, then there is also a confidently learnable $\eta$-family
which is not finitely learnable.
For all choices of $\eta$, the implications from confident to explanatory
learning and from vacillatory to behaviourally correct learning cannot
be reversed and the class of all
$\eta$-closed r.e.\ set is an $\eta$-family which cannot be learnt
behaviourally correctly.
The learnability of $\eta$-families with additional
constraints such as monotonicity and conservativeness
was also investigated; one interesting finding was that
the separability of learning criteria for $\eta$-families is sometimes
contingent on the existence of at least two $\eta$-infinite sets. 
\sp
Besides investigating the situation for further learning criteria,
future work can investigate to which extent the results generalise to arbitrary
uniformly r.e.\ families of $\eta$-closed sets. Here one would get that
the family of all $\eta$-singletons is r.e.\ and finitely, thus confidently
learnable: the learner generates an index of the $\eta$-equivalence
class to be learnt from the first data-item observed and keeps this
hypothesis forever. So one has one more level in the learning hierarchy.
However, the collapse of vacillatory learning to explanatory learning
for the constructed equivalence relation $\vartheta$ generalises to
uniformly r.e.\ families.  

\bibliographystyle{plain}

\end{document}